\documentclass[final]{siamltex}

\usepackage{latexsym}
\usepackage{amssymb}
\usepackage{amsfonts}
\usepackage{graphicx}
\usepackage{subfigure}
\usepackage{esint}

\def\eps{\varepsilon}

\def\sqgam{\sqrt{\gamma}}
\def\ga{\gamma}

\def\la{\lambda}
\def\La{\Lambda}

\def\th{\theta}
\def\tpsi{\widetilde{\psi}}
\def\O{{\cal O}}

\def\L{{\cal L}}

\def\pfl{\phi_{\rm flux}}
\def\pp3{\phi_{\pi}^3}

\def\gammacr{\gamma_{\rm cr}}

\newcommand{\beq}{\begin{equation}}
\newcommand{\eeq}{\end{equation}}
\newcommand{\sech}{\mbox{sech}}
\newcommand{\qmbox}[1]{\quad\mbox{#1}\quad}

\newtheorem{remark}[theorem]{remark}

\title{Stability analysis of $\pi$-kinks in a 0-$\pi$ Josephson junction\thanks{This work was supported by the Royal Netherlands Academy of Art and Sciences (KNAW) and Netherlands Organization for Scientific Research (NWO).}}

\author{G.\ Derks\thanks{Department of Mathematics, 
University of Surrey, Guildford, Surrey, GU2 7XH
({\tt g.derks@surrey.ac.uk}).}, 
\and
A.\ Doelman\thanks{Department 'Modelling, Analysis and Simulation', 
Center for Mathematics and Computer Science (CWI), 
Kruislaan 413, 1098 SJ Amsterdam, and 
Korteweg-de Vries Institute, Faculty of Sciences, 
University of Amsterdam, Plantage Muidergracht 24, 
1018 TV Amsterdam, The Netherlands ({\tt a.doelman@cwi.nl})},
\and
S.A.\ van Gils\thanks{Department of Applied Mathematics, 
University of Twente, P.O. Box 217, 7500AE Enschede, The Netherlands 
({\tt s.a.vangils@math.utwente.nl})},
\and
H.\ Susanto\thanks{Department of Mathematics and Statistics, 
University of Massachusetts, Amherst MA 01003-4515, USA 
({\tt susanto@math.umass.edu})}}

\begin{document}

\maketitle

\begin{abstract}
  We consider a spatially non-autonomous discrete sine-Gordon equation
  with constant forcing and its continuum limit(s) to model a 0-$\pi$
  Josephson junction with an applied bias current. The continuum
  limits correspond to the strong coupling limit of the discrete
  system. The non-autonomous character is due to the presence of a
  discontinuity point, namely a jump of~$\pi$ in the sine-Gordon
  phase. The continuum models admits static solitary waves which are
  called $\pi$-kinks and are attached to the discontinuity point. For
  small forcing, there are three types of $\pi$-kinks. We show that
  one of the kinks is stable and the others are unstable.  There is a
  critical value of the forcing beyond all static $\pi$-kinks fail to
  exist. Up to this value, the (in)stability of the $\pi$-kinks can be
  established analytically in the strong coupling limits. Applying a
  forcing above the critical value causes the nucleation of
  $2\pi$-kinks and -antikinks. Besides a $\pi$-kink, the unforced
  system also admits a static $3\pi$-kink. This state is unstable in
  the continuum models. By combining analytical and numerical methods
  in the discrete model, it is shown that the stable $\pi$-kink
  remains stable, and that the unstable $\pi$-kinks cannot be
  stabilized by decreasing the coupling. The $3\pi$-kink does become
  stable in the discrete model when the coupling is sufficiently weak.
\end{abstract}



\pagestyle{myheadings}
\thispagestyle{plain}
\markboth{DERKS ET AL.}{STABILITY ANALYSIS OF $\pi$-KINKS}

\section{Introduction}
\label{sec:intro}

One important application of the sine-Gordon equation is to describe
the propagation of magnetic flux (fluxons) in long Josephson junctions
\cite{kivs89,brau98}. The flux quanta or fluxons are described by the
kinks of the sine-Gordon equation. When many small Josephson junctions
are connected through the inductance of the superconductors, they form
a discrete Josephson transmission line. The propagation of a fluxon is
then described by the discrete sine-Gordon equation. For some
materials, Josephson junctions are more easily fabricated in the form
of a lattice than as a long continuous Josephson junction. In the
strong coupling limit, a discrete Josephson junction lattice becomes a
long Josephson junction.

It was proposed in the late 70's by Bulaevskii that a phase-shift of
$\pi$ may occur in the sine-Gordon equation due to magnetic
impurities~\cite{bula77}.  Only recently this prediction is confirmed
experimentally~\cite{vav06}.  Present technological advances can also
impose a $\pi$-phase-shift in a long Josephson junction using, e.g.,
superconductors with unconventional pairing symmetry~\cite{tsue00},
Superconductor-Ferromagnet-Superconductor~(SFS)
$\pi$-junctions~\cite{ryaz01}, or Superconductor-Normal
metal-Superconductor~(SNS) junctions in which the charge-carrier
population in the conduction channels is controlled~\cite{base99}. 

A junction containing a region with a phase jump of~$\pi$ is then
called a 0-$\pi$ Josephson junction and is described by a 0-$\pi$
sine-Gordon equation. The place where the 0-junction meets the
$\pi$-junction is called a discontinuity point.
A 0-$\pi$ Josephson junction admits a half magnetic flux (semifluxon),
sometimes called \emph{$\pi$-fluxon}, attached to the discontinuity
point~\cite{hilg03}. A semifluxon is represented by a $\pi$-kink in
the 0-$\pi$ sine-Gordon equation~\cite{susa03}. 

Using the technology described in~\cite{hilg03}, a 0-$\pi$ array of
Josephson junctions can be created as well. Such system can be modeled
by a discrete $0$-$\pi$ sine-Gordon equation. A short numerical study of
a discrete $\pi$-kink is given in~\cite{susa04}. 

The presence of the semifluxon in a 0-$\pi$ Josephson junction or a
0-$\pi$ array of Josephson junctions opens a new field where many
questions, that have been discussed in detail for the $2\pi$-kink
(fluxon) in the sine-Gordon equation, can be addressed for the
$\pi$-kink too. The fact that the $\pi$-kink cannot move in space,
even in the continuum case, will give a different qualitative behavior
such as the disappearance of the zero eigenvalue (Goldstone mode), as
will be shown later.

In this paper we will study both the continuous and discrete 0-$\pi$
sine-Gordon equation, especially the stability of the kinks admitted
by the equations. Knowing the eigenvalues of a kink is of interest for
experimentalists, since the corresponding eigenfunctions (localized
modes) can play an important role in the behavior of the
kink~\cite{quin00}. 

The present work is organized as follows: 
in section~\ref{3sec2} we will describe the mathematical model of the
problem and its interpretation as a Josephson junction system. We will
discuss the discrete system as well as several continuum
approximations.
In section~\ref{3sec3} we consider the continuous 0-$\pi$ sine-Gordon
equation which describes a continuous long Josephson junction with
discontinuity point. It is also the lowest order continuum
approximation for the discrete system, not reflecting any lattice spacing
(coupling) effects. In~\cite{susa03} it is shown that there
exist three types of $\pi$-kinks in the 0-$\pi$ sine-Gordon equation.
We will analyze their stability and show that one type is stable and
the other two are unstable.
A higher order continuum approximation, which includes terms
representing a small lattice spacing (strong coupling), is considered in
section~\ref{3sec4} . It is shown that for small values of the lattice
spacing parameter, the three types of $\pi$-kinks persist and their stability
properties do not change.
In section~\ref{3sec5} the discrete 0-$\pi$ sine-Gordon with large
lattice spacing (small coupling) is analyzed, especially the existence
and stability of $\pi$-kinks. 
Numerical calculations connecting the regions of small and large
lattice spacing (weak and strong coupling) will be presented in
section~\ref{3sec6}. In this section the analytical results
of the previous sections are linked together.
Conclusions and plans for future research are presented in
section~\ref{3sec7}.

\section{Mathematical models for 0-$\pi$ junctions}
\label{3sec2}

\subsection{The discrete 0-$\pi$ sine-Gordon equation}

The Lagrangian describing the phase of a $0$-$\pi$ array of Josephson
junctions is given by
\begin{equation}
L=\int
{\sum_{n\in\mathbb{Z}}{\left[\frac12\left(\frac{d\phi_n}{dt} \right)^2
-\frac12\left(\frac{\phi_{n+1}-\phi_n}{a}\right)^2
-1+\cos(\phi_n+\theta_n)+\gamma\phi_n\right]}\, dt},
\label{hamilton}
\end{equation}
where $\phi_n$ is the Josephson phase of the $n$th junction. The phase
jump of $\pi$ in the Josephson phase is described by $\theta_n$ where
\begin{equation}
\theta_n=\left\{\begin{array}{rl}
0,&n\leq0,\\
-\pi,&0< n.
\end{array}\right.
\label{theta}
\end{equation}

The Lagrangian~(\ref{hamilton}) is given in nondimensionalized form.
The lattice spacing parameter~$a$ is normalized to the Josephson
length~$\lambda_J$, the time~$t$ is normalized to the inverse plasma
frequency~$\omega_0^{-1}$ and the applied bias current
density~$\gamma>0$ is scaled to the critical current density~$J_c$.

The equation of the phase motion generated by the
Lagrangian~(\ref{hamilton}) is the discrete $0$-$\pi$
sine-Gordon equation
\begin{equation}
\ddot{\phi_n} - \frac{\phi_{n-1}-2\phi_n+\phi_{n+1}}{a^2} =
-\sin(\phi_n+\theta_n)+\gamma.
\label{dsg}
\end{equation}

We use $n\in\mathbb{Z}$ for the analytical calculations, but of
course, the fabrication of the junction as well as the numerics are
limited to a finite number of sites, say~$2N$. We will take the
boundary conditions to represent the way in which the applied magnetic
field $h=H/(\lambda_JJ_c)$ enters the system, i.e., 
\begin{equation}
\frac{\phi_{-N+1}-\phi_{-N}}{a}=\frac{\phi_{N}-\phi_{N-1}}{a}=h.
\label{bc_dsg}
\end{equation}
In the sequel we will always consider the case when there is no applied
magnetic field, i.e., we will take $h=0$.

\subsection{Approximations to the lattice spacing in the continuum
  limit}

There are various continuum model approximations for~(\ref{dsg}) that
can be derived in the continuum limit $a\ll 1$.
Writing $\phi_n=\phi(na)$ and expanding the difference terms using a
Taylor expansion gives
\[
\frac{\phi_{n-1}-2\phi_n+\phi_{n+1}}{a^2} 
= 2\sum_{k=0}^\infty \frac{a^{2k}}{(2k+2)!} \, \partial^{k}_{xx} 
\phi_{xx}(na) = L_a\phi_{xx}
\]
and
\[
\frac{\phi_{n+1}-\phi_n}{a} = 
\sum_{k=0}^\infty \frac{a^{k}}{(k+1)!} \, \partial^k_x \phi(na) = 
\widetilde L_a\phi_x.
\]
Thus the continuum approximation for~(\ref{dsg})  is 
 \begin{equation}\label{eq.cont_full}
\phi_{tt} - L_a\phi_{xx} = -\sin(\phi+\theta)+\gamma, 
\end{equation}
where $\theta(x)$ is defined similar to~(\ref{theta}), i.e.,
\[
\theta(x)=\left\{\begin{array}{ll}
0,&x<0,\\
-\pi,&x> 0.
\end{array}\right.
\]

The continuum approximation for the Lagrangian is 
\[
L=\iint_{-\infty}^\infty
\left[ \,\frac{1}{2}\left(\phi_t \right)^2
-\frac{1}{2}\left(\widetilde L_a \phi_x\right)^2
-1+\cos(\phi+\theta)+\gamma\phi\right]\, dx\,dt\,
\]
Note that the normalizations in the discrete system imply that the
spatial coordinate~$x$ is normalized to the Josephson length~$\lambda_J$.

There are several ways to derive approximations for the operator~$L_a$
when $a\to0$, see for example~\cite{rose03}.  The
first obvious approximation is
\begin{equation}
  \label{eq.mod1}
\phi_{tt} - \phi_{xx} - \frac{a^2}{12}\phi_{xxxx} = 
-\sin(\phi+\theta)+\gamma, \quad x\neq 0.
\end{equation}
%
Another approximation can be found by using that 
$(1-\frac{a^2}{12}\partial_{xx})\,L_a = 1-
\frac{a^4}{240}\partial^2_{xx}+\ldots$. This results reflects the
invertibility of $L_a$ up to fourth order. Hence
$(1-\frac{a^2}{12}\partial_{xx})$ acting on~(\ref{eq.cont_full}) gives
the approximation (up to fourth order terms) 
\begin{equation}
\label{eq.mod2}
\phi_{xx} = \phi_{tt} +\sin(\phi+\theta)-\gamma - 
\frac{a^2}{12}\partial_{xx}(\phi_{tt}
+\sin(\phi+\theta)), \quad x\neq 0 .
\end{equation}
Expanding this equation and using the expression for $\phi_{xx}$
again, we get
\begin{eqnarray}
\label{eq.mod4}
\phi_{xx} &=& \phi_{tt} +\sin(\phi+\theta)-\gamma \\
&&-\frac{a^2}{12}\left(\phi_{tttt} +[\sin(\phi+\theta)]_{tt}-\phi^2_x\sin(\phi+\theta)\right.\nonumber\\
&&\left.{}+\cos(\phi+\theta)[\phi_{tt}+\sin(\phi+\theta)-\gamma] \right), \quad x\neq 0 .\nonumber
\end{eqnarray}

The steady state equation for~(\ref{eq.mod1}) is
\[
\phi_{xx} + \frac{a^2}{12}\phi_{xxxx} = 
\sin(\phi+\theta)
-\gamma, \quad x\neq 0 ,
\]
while~(\ref{eq.mod2}) yields the equation
\[
\phi_{xx} = 
(1-\frac{a^2}{12}\partial_{xx})\sin(\phi+\theta)
-\gamma, \quad x\neq 0 ,
\]
and~(\ref{eq.mod4}) gives
\[
\phi_{xx} =\sin(\phi+\theta)-\gamma - 
\frac{a^2}{12}(
- \phi^2_x\sin(\phi+\theta)
+\cos(\phi+\theta)[\sin(\phi+\theta)-\gamma]
), \quad x\neq 0 .
\]
Unfortunately the last two equations are not Hamiltonian, so we have
lost the Hamiltonian properties of the original system, while
the first equation is singularly perturbed.

Yet another approximation that has a variational structure and is not
singularly perturbed can be obtained by combining the two equations
that have lost their variational character. Indeed,
taking~(\ref{eq.mod2}) twice and subtracting~(\ref{eq.mod4}) gives
\begin{eqnarray}
  \label{cont_appr}
\hspace{1cm}\phi_{xx} &=& \phi_{tt} +\sin(\phi+\theta)-\gamma \\
&&{}-\frac{a^2}{12}
\left( 2\phi_{xxtt}+2\phi_{xx}\cos(\phi+\theta)-\phi_x^2\sin(\phi+\theta)-\phi_{tttt} -\phi_{tt}\cos(\phi+\theta)\right.\nonumber\\
&&\left.{}+\phi_t^2\sin(\phi+\theta)-\cos(\phi+\theta)(\phi_{tt}+\sin(\phi+\theta)-\gamma)\right), \quad x\neq 0.\nonumber  
\end{eqnarray}
The Lagrangian for this system is 
\begin{eqnarray}
L&=&
\displaystyle\iint\textstyle
\frac12\phi_t^2-\frac12\phi_x^2-1+\cos(\phi+\theta)+\gamma\phi\nonumber\\
&&\hspace{1cm}+\frac{a^2}2\left[ \phi_x\partial_x(\phi_{tt}+\sin(\phi+\theta))+\frac12(\phi_{tt}+\sin(\phi+\theta)-\gamma)^2 \right]
\, dx\,dt\,.\nonumber
\end{eqnarray}
The static equation for~(\ref{cont_appr}) is
\begin{eqnarray}
\label{cont_appr_static}
\hspace{1cm}
&&\phi_{xx} = \sin(\phi+\theta)-\gamma \\
&&{}-\frac{a^2}{12}\left(
2\phi_{xx}\cos(\phi+\theta) - \phi^2_x\sin(\phi+\theta)-\cos(\phi+\theta)(\sin(\phi+\theta)-\gamma)\right), \quad x\neq 0. \nonumber
\end{eqnarray}
This equation is a regularly perturbed Hamiltonian system with the
Hamiltonian
\[
H(\phi,p)= \frac{p^2}{2(1+\frac{a^2}6\cos(\phi+\theta))} + \gamma\phi
+ \cos(\phi+\theta) - \frac{a^2}{24}\left(\sin(\phi+\theta)-\gamma
\right)^2,
\]
which implies $p=\phi_x\left(1+\frac{a^2\cos(\phi+\theta)}6\right)$. 

In this paper, we will analyze equation~(\ref{cont_appr}) as a
continuum strong interaction limit which incorporates some effects of
the lattice spacing into the model. The model
equation~(\ref{cont_appr}) is chosen as it is non-singular and has
the same conservative properties as the discrete system, reflecting
its physical properties.

\section{The $\pi$-kinks and their spectra in the continuum limit}
\label{3sec3}

In this section, we will consider~(\ref{cont_appr}) for $a=0$, which
is a model for an ideal long 0-$\pi$ Josephson junction:
\begin{equation}\label{eq:pSG}
  \phi_{tt}-\phi_{xx} + \sin(\phi+\theta)= \gamma,
\quad x \neq 0.
\end{equation}
For a Josephson junction without an applied bias current or a phase jump,
i.e., for ~$\ga=0$ and $\th(x)\equiv0$, the model corresponds to
the sine-Gordon equation.
A stable solution of the sine-Gordon equation is the basic
(normalized) stationary, monotonically increasing fluxon, given by
\beq 
\label{eq:funflux}
\pfl(x) = 4 \arctan e^x, \quad \pfl(0) = \pi
\eeq
(see \cite{derk03}). 

In general the discontinuous function $\theta(x)$ in~(\ref{eq:pSG})
will introduce a discontinuity at $x=0$ for the second derivative
$\phi_{xx}$. Hence, the natural solution space for~(\ref{eq:pSG})
consists of functions which are spatially continuous and have a
continuous spatial derivative. The behavior at infinity is regulated
by requiring that the spatial derivative of the solution belongs to
$H_1(\mathbb{R})$ (which allows the phase to converge to a nonzero
constant at infinity). Therefore, the equation~(\ref{eq:pSG}) is
considered as a dynamical system on the function space
\[
\mathbb{H} = \{ \phi:\mathbb{R}\to\mathbb{R} \mid \phi_x \in
H_1(\mathbb{R}) \}.
\]
 
It is straightforward to find that for $|\gamma|<1$ and $x<0$, the
``fixed points'' of~(\ref{eq:pSG}) are $\phi_s^-=\arcsin(\gamma)$ and
$\phi_c^-=-\arcsin(\gamma)+ \pi$. Similarly, for $|\gamma|<1$ and
$x>0$, they are $\phi_s^+ = \arcsin(\gamma)+ \pi$ and $\phi_c^+ =
-\arcsin(\gamma)+ 2\pi$. In~\cite{susa03}, it is shown that there
exists various types of stationary fronts, which connect the
equilibria. Most stationary fronts are so-called $\pi$-kinks, which are
static waves connecting equilibrium states at $x=\pm\infty$ with a
phase-difference of~$\pi$. Such waves are solutions of the static wave
equation 
\begin{equation}
  \label{eq:static}
  \phi_{xx} - \sin(\phi+\theta) = -\gamma  , \quad x\neq0.  
\end{equation}
In the $x$-dynamics of~(\ref{eq:static}), the points $\phi_s^\pm$ are
saddle points and the points with $\phi_c^\pm$ are center points. Thus
a $\pi$-kink connects $\phi_s^-$ with $\phi_s^+$. 

In this section we will consider the stability of those $\pi$-kinks.
For completeness, we first describe the various types of $\pi$-kinks
as found in~\cite{susa03}. These $\pi$-kinks are constructed by taking
suitable combinations of the phase portraits for $\theta=0$ and
$\theta=-\pi$. The phase portraits for $\gamma=0$ are essentially
different from the ones for $0<\gamma<1$ (the case $-1<\gamma<0$
follows from this one by taking $\phi\mapsto -\phi$ and
$\gamma\mapsto-\gamma$).  In case $\gamma>0$ there are homoclinic
connections at $k\pi+\arcsin(\gamma)$, $k\in{\mathbb Z}$, $k$ even
($\theta=0$) or $k$ odd ($\theta=-\pi$).  If $\gamma=0$, then these
homoclinic connections break to heteroclinic connections between
$k\pi$ and $(k+2)\pi$.

The phase portrait of~(\ref{eq:static}) for $\gamma=0$ is shown in
Figure~\ref{g0}(a).  Following the notation in~\cite{susa03}, in case
$\gamma=0$, there are two types of heteroclinic connections (kinks) in
the 0-$\pi$ junction. The first one, called \emph{type~1}
and denoted by $\phi_\pi^1(x;0)$, connects~$0$ and~$\pi$.  The point
in the phase plane where the junction lies is denoted by $d_1(0)$.
The second one, called \emph{type~2} and denoted by
$\phi_{3\pi}^2(x;0)$, connects~$0$ and~$3\pi$. Now the point in the
phase plane where the junction lies is denoted by $d_2(0)$.  This
solution is not a semifluxon, but it will play a role in the analysis
of some of the semifluxons for $\gamma\neq0$.

\begin{figure}[tbp!]
\begin{center}
\vspace{2cm}
\end{center}
\caption{(a)The
  phase portrait of system~(\ref{eq:static}) for $\gamma=0$. The
  trajectories for $x<0$ are indicated with bold lines, the
  trajectories for $x>0$ with dashed lines. Any orbit
  of~(\ref{eq:pSG}) switches at $x=0$ from bold to dashed. The type~1
  semifluxon switches at $d_1$ and corresponds to one of the gray arrow-lines.
  The $3\pi$-fluxon switches at $d_2$ and is denoted by the other gray
  arrow-line.\newline 
(b) The phase portrait of system
  (\ref{eq:static}) for $\gamma=0.1$. For simplicity, only the stable
  and unstable manifolds of the fixed points are shown. Apart from
  $d_1$, there are also the points $d_2$ or $d_3$ which can be used
  for the switch position of $x=0$ to obtain a solution with a phase
  difference $\pi$ between the end points.}
\label{g0}
\end{figure}

If $0<\gamma\ll 1$, then there are three types of $\pi$-kinks
(heteroclinic connections) in the junction, all
connecting~$\arcsin(\gamma)$ and $\pi+\arcsin(\gamma)$.  A phase
portrait of~(\ref{eq:static}) for nonzero $\gamma$ is shown in
Figure~\ref{g0}(b).  The first semifluxon, called \emph{type~1} and
denoted by $\phi_\pi^1(x;\gamma)$, is a continuation of the connection
at $\gamma=0$.  The point in the phase plane where the junction lies
is denoted by $d_1(\gamma)$. The $\pi$-fluxon~$\phi_\pi^1(x;\gamma)$
is monotonically increasing.

The second one is called \emph{type~2} and denoted by
$\phi_\pi^2(x;\gamma)$. In the limit for $\gamma\to0$, it breaks in
the $3\pi$-kink and the heteroclinic connection between $3\pi$ and
$\pi$ (a $-2\pi$-kink or an anti-fluxon). The point in the phase plane
where the junction lies is denoted by $d_2(\gamma)$. The
$\pi$-fluxon~$\phi_{\pi}^2(x;\gamma)$ is not monotonically increasing,
but has a hump.

The third one is called \emph{type~3} and denoted by
$\phi_\pi^3(x;\gamma)$. In the limit for $\gamma\to0$, it breaks in
the heteroclinic connection between $0$ and $2\pi$ (fluxon) and an
anti-semifluxon like the type~1 wave but connecting $2\pi$ and $\pi$.
The point in the phase plane where the junction lies is denoted by
$d_3(\gamma)$. This $\pi$-fluxon has a hump too, but a lower one than
the type~2 wave.  Following the first homoclinic orbit, the junction
points are ordered such that $d_1(\gamma)$ comes first, followed by
$d_2(\gamma)$, followed by $d_3(\gamma)$ (see Figure~\ref{g0}(b)).

If $\gamma$ increases, the points $d_2(\gamma)$ and $d_3(\gamma)$
approach each other, until they coincide at \cite{susa03}
\begin{equation}
\gamma=\gamma^*=\frac{2}{\sqrt{4+\pi^2}} 
\label{gammastar}
\end{equation}
in the point $(\pi+\arcsin(\gamma^*),0)$. At this point, the type~2
wave $\phi_\pi^2(x;\gamma)$ ceases to exist (in the limit it breaks
into half the homoclinic connection for $x<0$ and the full homoclinic
connection for $x>0$). The type~3 kink $\phi_\pi^3(x;\gamma^*)$
consists of half the homoclinic connection for $x<0$ and the fixed
point for $x>0$ and this wave can be continued for
$\gamma>\gamma^*$. For $\gamma>\gamma^*$, the type~3 kink is
monotonic. 

If $\gamma$ increases further, the points $d_1(\gamma)$ and
$d_3(\gamma)$ approach each other~\cite{susa03} until they coincide at
\begin{equation}
\gamma=\gammacr=\frac2\pi.
\label{gammacr}
\end{equation}
When $\gamma=\gammacr$, the orbit homoclinic to the hyperbolic fixed
point for $x<0$ is tangential at $d_1(\gamma)=d_3(\gamma)$ to the
non-homoclinic stable manifold of the hyperbolic fixed point for $x>0$.
As soon as $\gamma>\gammacr$, there is no more intersection of the
homoclinic orbit for $x<0$ with a stable manifold of the hyperbolic
fixed point for $x>0$. This implies that no static $\pi$-fluxons can
exist.  For more details, see~\cite{susa03}.

After recalling the description of the $\pi$-kinks from~\cite{susa03},
we can start the stability analysis. It will be shown that the
type~1 $\pi$-kink is nonlinearly stable for all $0\leq \gamma \leq
\gammacr$. The type~2 and type~3 $\pi$-kinks are linearly unstable for
all values of $\gamma$ for which they exist. First we consider the
linearization about the $\pi$-kinks.
\begin{theorem}
  \label{th:stab1}
The linearizations about the various $\pi$-kinks have the following
properties:
\begin{romannum}
\item The eigenvalues of the linearization about the monotonic type~1
  $\pi$-kink $\phi_\pi^1(x;\gamma)$ are strictly negative for
  $0\leq\gamma<\gammacr$. At $\gamma=\gammacr$, the largest eigenvalue is
  zero. These $\pi$-kinks are linearly stable.
\item The largest eigenvalue of the linearization about the 
  type~2 $\pi$-kink $\phi_\pi^2(x;\gamma)$ is strictly positive for
  $0<\gamma<\gamma^*$. These $\pi$-kinks are linearly unstable.
\item The largest eigenvalue of the linearization about the 
  type~3 $\pi$-kink $\phi_\pi^3(x;\gamma)$ is strictly positive for
  $0<\gamma<\gammacr$. These $\pi$-kinks are linearly unstable. In the
  limit for $\gamma\to0$ and $\gamma\to\gammacr$, the largest
  eigenvalue converges to zero.
\end{romannum}
\end{theorem}

\begin{remark} \rm Note that the instability of the two non-monotonic
  $\pi$-kinks cannot be established by the classical Sturm-Liouville
  argument.  In the classical, autonomous setting, the derivative of
  the wave about which the system is linearized, is an eigenfunction
  of the linearized system. This eigenfunction is associated with the
  translation invariance of the original system and hence corresponds
  to an eigenvalue $\lambda = 0$. If the wave is non-monotonic, then
  its derivative has a zero, which implies that $\lambda = 0$ is not
  the largest eigenvalue~\cite{Titch} and that the wave must be
  unstable.  Due to the discontinuity at $x = 0$, our system is
  non-autonomous, thus not invariant with respect to translations, and
  $\lambda = 0$ is (in general) not an eigenvalue.  Thus, it cannot a
  priori be concluded that the non-monotonic $\pi$-kinks must be
  unstable.
\end{remark}

To prove Theorem~\ref{th:stab1}, it will be shown that the
linearization about a $\pi$-kink has an eigenvalue zero if and only if
the $\pi$-kink takes a value which is a multiple of $\pi$ at $x=0$.
Since the value at $x=0$ is related to the point $d_i(\gamma)$, it can
be derived that this happens only at $\gamma=\gammacr$ for the
colliding type~1 and type~3 waves.  To complete the proof, we will
derive expressions for the largest eigenvalue of the linearization
about each semikink near $\gamma=0$ in three separate lemmas and use
that the eigenvalues are continuous in~$\gamma$ to derive the sign of
the largest eigenvalue on the existence interval of the $\pi$-kink.

To linearize about a solution $\phi^i_\pi(x;\gamma)$, write
$\phi(x,t)= \phi^i_\pi(x;\gamma) + v(x,t)$, substitute this in the
model equation~(\ref{eq:pSG}) and disregard all higher order terms:
\begin{equation}
  \label{eq:linsG}
[D_{xx} - \cos(\phi^i_\pi(x;\gamma)+\theta(x))]  \, v 
= D_{tt}\,v.
\end{equation}
Using the spectral Ansatz $v(x,t)=e^{\lambda t} \widetilde v(x)$,
where $v(x)$ is a continuously differentiable function and
dropping the tildes, we get the eigenvalue problem
\begin{equation}\label{eq:eigenval}
{\cal L}^i(x;\gamma) \, v = \lambda^2\, v,
\end{equation}
where ${\cal L}^i$ is defined as
\begin{equation}
  \label{eq:lin}
{\cal L}^i(x;\gamma) = D_{xx} -\cos(\phi^i_\pi(x;\gamma)+\theta(x)) . 
\end{equation}
The natural domain for ${\cal L}^i$ is $H_2(\mathbb{R})$. 
We call $\Lambda$ an eigenvalue of ${\cal L}^i$ if there is a function
$v\in H_2(\mathbb{R})$,  which satisfies 
${\cal L}^i(x;\gamma) \, v =\Lambda v$.
Since ${\cal L}^i$ depends smoothly on~$\gamma$,
the eigenvalues of ${\cal L}^i$ will depend smoothly on~$\gamma$
too. 
 
The operator ${\cal L}^i$ is symmetric, hence all eigenvalues will be
real. A straightforward calculation gives that the continuous spectrum
of ${\cal L}^i$ is in $(-\infty,-\sqrt{1-\gamma^2})$.
Since the eigenfunctions are continuously differentiable functions in
$H_2(\mathbb{R})$ by the Sobolev Embedding Theorem, Sturm's Theorem~\cite{Titch}
can be applied, leading to the fact that the eigenvalues are bounded
from above.  Furthermore, if $v_1$ is an eigenfunction of ${\cal L}^i$
with eigenvalue $\Lambda_1$ and $v_2$ is an eigenfunction of ${\cal
  L}^i$ with eigenvalue $\Lambda_2$ with $\Lambda_1>\Lambda_2$, then
there is at least one zero of $v_2$ between any pair of zeros of $v_1$
(including the zeros at $\pm\infty$). Hence if the eigenfunction $v_1$
has fixed sign, then $\Lambda_1$ is the largest eigenvalue of ${\cal
  L}^i$.

The following lemma gives a necessary and sufficient condition for
${\cal L}^i$ to have an eigenvalue $\Lambda=0$. 
\begin{lemma}\label{lem:zero}
The eigenvalue problem 
\[
{\cal L}^i(x;\gamma) v = \Lambda v, \quad x\in\mathbb{R},
\]
has an eigenvalue $\Lambda =0$ if and only if one of the following two
conditions holds 
\begin{romannum}
\item $D_{xx}\phi^i_\pi(x;\gamma)$ is continuous at $x=0$, i.e.,
  $\phi^i_\pi(0;\gamma)=k\pi$, for some $k\in \mathbb{Z}$;
\item $D_x\phi_\pi^i(0;\gamma) =0$ and there are some $x_\pm$, with
  $\mathrm{sgn}(x_\pm)= \pm1$, such that $D_x\phi_\pi^i(x_\pm;\gamma)\neq 0$.
\end{romannum}
\end{lemma}
\begin{proof}
Since $\phi_\pi^i(x;\gamma)$ converges to a saddle point for $|x|\to\infty$,
this implies that $D_x\phi_\pi(x;\gamma)$ decays exponentially fast to 0 for 
$|x|\to\infty$. 
Since $\phi_\pi^i(x;\gamma)$ solves~(\ref{eq:static}), differentiating this
ODE with respect to $x$, gives
\[
{\cal L}^i (x;\gamma) \, D_x\phi_\pi^i (x;\gamma) = 0, \qmbox{for} x\neq 0.  
\]
This implies that for any constant $K$, the function $w^i_K(x)=
K\,D_x\phi_\pi^i (x;\gamma)$ satisfies ${\cal L}^i (x;\gamma) \, w^i_K(x) =0$
for $x\neq 0$. Hence for any $K_-$ and $K_+$, the solution
\[
w^i(x) = \left\{
\begin{array}{ll}
  w^i_{K_-}(x), & x<0,\\
w^i_{K_+}(x), & x>0,
\end{array}\right.
\]
solves ${\cal L}^i (x;\gamma) \, w^i(x) =0$
for $x\neq 0$. The function $w^i(x)$ is continuously differentiable if
and only if the following two conditions hold
\begin{enumerate}
\item $w^i_{K_-}(0-)=w^i_{K_+}(0+)$, in other words,
$K_-\, D_x\phi^i_\pi (0;\gamma ) = K_+\, D_x\phi_\pi^i (0;\gamma )$, since
$\phi_\pi^i$ is continuously differentiable; 
\item
$D_xw^i_{K_-}(0-)=D_xw^i_{K_+}(0+)$, thus
$K_-\, D_{xx}\phi_\pi^i (0-;\gamma) = K_+\, D_{xx}\phi_\pi^i (0+;\gamma)$.
\end{enumerate} 
The first condition is satisfied if $K_-=K_+$ or $D_x\phi_\pi^i
(0;\gamma)=0$. If $D_x\phi_\pi^i(0;\gamma)=0$, we can choose $K_\pm$
such that the second condition is satisfied and we do not end up with
the trivial solution, except when $D_{x}\phi_\pi^i (x;\gamma)$ is trivial for
either $x>0$ or $x<0$. 

If $D_x\phi_\pi^i(0;\gamma)\neq 0$, we need $D_{xx}\phi_\pi^i$ to be
continuous at $x=0$ in order to satisfy the second condition. Since
$D_{xx}\phi^i_\pi(x;\gamma)=\sin(\phi_\pi^i(x;\gamma)+\theta(x)) -\gamma$,
$D_{xx}\phi^i_\pi$ is continuous at $x=0$ if and only if
$\sin(\phi_\pi^i(0;\gamma))=0$. 
These arguments prove that if one of the two conditions are satisfied,
then $\Lambda=0$ is an eigenvalue of ${\cal L}^i$. 

Next we assume that $\Lambda=0$ is an eigenvalue of ${\cal L}^i$,
hence there is some continuously differentiable function $v^i(x)$ 
such that ${\cal L}^i(x)v^i(x)=0$ for $x\neq 0$ and $v^i(x)\to0$
for $|x|\to\infty$. The only solutions decaying to zero at $+\infty$
are the solutions on the one-dimensional stable manifold and
similarly, the only solutions decaying to zero at $-\infty$
are the solutions on the one-dimensional unstable manifold. The stable
and unstable manifold are formed by multiples of $D_x\phi_\pi^i$. So we
can conclude that there exist $K_\pm$ such that
\[
v^i(x) = \left \{ 
  \begin{array}{lll}
    K_-D_x\phi_\pi^i(x)&\qmbox{for}&x<0,\\
    K_+D_x\phi_\pi^i(x)&\qmbox{for}&x>0.
  \end{array}\right.
\] 
Now we are back in the same situation as above, so we can conclude
that either one of the two conditions in the lemma must be satisfied.
\end{proof}

The second condition in the lemma does not occur.  Indeed, the first
part of the second condition, i.e., $D_x\phi_\pi^i(0;\gamma)=0$
happens only if $d_i$ has its second coordinate zero, hence only at
$\gamma=\gamma^*$ with $d_2=d_3$. At this point, the solution
$\phi_\pi^2(x;\gamma^*)$ has ceased to exist and the solution
$\phi_\pi^3(x;\gamma^*)$ consists of the fixed point for $x>0$. Hence
this solution does not satisfy the second part of the second
condition.

To see for which value of $\gamma$ the first condition is satisfied, we
derive the relation between $\phi_\pi^i(0;\gamma)$ and $\gamma$. 
Multiplying the static equation~(\ref{eq:static}) with $D_x\phi^i_\pi$
and rewriting it gives
\[
D_x[(D_x\phi_\pi^i(x;\gamma))^2] = 2 D_x [ -\gamma \phi_\pi^i(x;\gamma)
- \cos(\phi_\pi^i(x;\gamma)+\theta(x)) ], \quad x\neq 0. 
\]
Integration from $\pm\infty$ to 0 and using that
$D_x\phi_\pi^i(\pm\infty;\gamma)=0$, shows
\begin{eqnarray}
(D_x\phi_\pi^i(0;\gamma))^2 &=& 2[
-\gamma (\phi_\pi^i(0;\gamma)-\phi_\pi^i(-\infty;\gamma))
- \cos(\phi_\pi^i(0;\gamma))+\cos(\phi_\pi^i(-\infty;\gamma))],\nonumber\\
(D_x\phi_\pi^i(0;\gamma))^2 &=& 2[
-\gamma (\phi_\pi^i(0;\gamma)-\phi_\pi^i(+\infty;\gamma))
+ \cos(\phi_\pi^i(0;\gamma))-\cos(\phi_\pi^i(+\infty;\gamma))].\nonumber
\end{eqnarray}
Subtracting these two equations and using that
$\phi_\pi^i(+\infty;\gamma)= \phi_\pi^i(-\infty;\gamma)+\pi$, we get
that
\begin{equation}\label{eq:phi0}
0=-\pi\gamma -2\cos(\phi_\pi^i(0;\gamma)), \qmbox{hence}
\cos(\phi_\pi^i(0;\gamma)) = \frac{\pi\gamma}{2}.
\end{equation}
Thus the first condition is only satisfied when
$\cos(\phi_\pi^i(0;\gamma))=\pm1$, hence
$\gamma=\frac{2}{\pi}=\gammacr$. 
%

The following step in the analysis of the eigenvalues of the
linearization is to consider the behavior of the eigenvalues for $\gamma$
small. First note that at $\gamma=0$, we have an explicit expression
for the $\pi$-fluxon and the $3\pi$-fluxon (see~(\ref{eq:funflux}) for
the expression of $\pfl$):
\begin{eqnarray}
  \phi^1_\pi(x;0) &=& \left\{
    \begin{array}{ll}
      \pfl(x-\ln(1+\sqrt2)),&\qmbox{for}\,x<0,\\
      \pi-\pfl(-x-\ln(1+\sqrt2)),&\qmbox{for}\,x>0,
    \end{array}\right.\label{sge2a}\\
      \phi^2_{3\pi}(x;0) &=& \left\{
    \begin{array}{ll}
      \pfl(x+\ln(1+\sqrt2)),&\qmbox{for}\,x<0,\\
      3\pi-\pfl(-x+\ln(1+\sqrt2)),&\qmbox{for}\,x>0.
    \end{array}\right.\label{sge2b}
\end{eqnarray}
Hence the derivatives of both functions are even and
$\cos(\phi^i_\pi(x;0)+\theta)$ is continuous and even, since
$\phi^1_\pi(0;0)=\frac{\pi}2$ and $\phi^2_{3\pi}(0;0)=\frac{3\pi}2$.

For $\gamma\ll 1$, the homoclinic orbit in the system with $\theta=0$
will be crucial for the approximation of type~2 and type~3
solutions. This orbit is homoclinic to $\arcsin(\gamma)$ and will be
denoted by $\phi_h(x;\gamma)$. 
It can be approximated up to order $\gamma$ by
using the $2\pi$-fluxon $\pfl$ and its linearization.
\begin{lemma}\label{lem:hom}
For $\gamma$ small, we have for the even homoclinic connection
$\phi_h(x;\gamma)$ 
\beq
\label{eq:expphi}
\hspace{1cm}\phi_h(x;\gamma) = \pfl(x+L_\pi(\gamma)) + \gamma
\,\phi_1(x+L_\pi(\gamma)) + \gamma^2 R_2(x+L_\pi(\gamma);\gamma),\,x<0, 
\eeq 
where the expression for the $2\pi$-fluxon $\pfl$ can be found
in~(\ref{eq:funflux}),
\[
\phi_1(x) = \frac12 \,\left[-1 + \cosh x + \int_0^x \frac{\xi}{\cosh
    \xi} d \xi\right] \,\frac{1}{\cosh x} - \arctan
e^x\,\left(\frac{x}{\cosh x} + \sinh x\right)
\]
and
$L_\pi(\gamma)$ is such
that 
$\phi_h(-L_\pi(\gamma);\gamma)=\pi=\pfl(0)$, implying 
\begin{equation}
L_\pi(\gamma) = \frac12 |\ln \gamma| + \ln \frac{4}{\sqrt{\pi}} +
\O(\sqrt\gamma).\label{Lpi}
\end{equation}
Furthermore, $\gamma^2R_2(x+L_\pi(\gamma);\gamma) = {\cal O}(\gamma)$,
uniform for $x<0$ 
and $\gamma\phi_1(L_\pi(\gamma);\gamma)={\cal O}(\sqrt\gamma)$. Thus
\beq
\label{pp3Lpi}
\phi_h(0) = 2 \pi - 2 \sqrt{\pi} \sqrt{\gamma} + \O(\gamma).
\eeq
Finally,
$\phi_1(\tilde x;\gamma)= {\cal O}(1)$ and 
$R_2(\tilde x;\gamma) = {\cal O}(1)$, uniform for
$\tilde x<0$.
\end{lemma}

\begin{proof}
  It is more convenient in the following perturbation analysis to
  follow the normalization of $\pfl(x)$,
  i.e., in this proof we introduce new coordinates $\tilde x =
  x+L_\pi(\gamma)$ where $L_\pi(\gamma)$ is such that
  $\phi_h(-L_\pi(\gamma);\gamma) = \pi=\pfl(0)$. In the following we
  will drop the tildes and work in those new coordinates.
  As $\phi_h$ in the original coordinates was even, we get in the new
  coordinates $D_x\phi_h(L_\pi(\gamma);\gamma)=0$. This condition will
  be used later to determine an asymptotic expression for
  $L_\pi(\gamma)$.

In the new coordinates, we introduce the expansion
\[
\phi_h(x; \gamma) = \pfl(x) + \gamma
\phi_1(x) + \gamma^2 R_2(x; \gamma) ,\quad x< L_\pi(\gamma).  
\]
By linearizing about $\pfl$, it follows that the equation for $\phi_1$ is
\beq
\label{eq:phi1}
{\cal L}(x)\,\phi_1 = - 1, \qmbox{where} 
{\cal L}(x) = D_{xx} - \cos (\pfl(x)).
\eeq
The operator ${\cal L}(x)$ is identical to the operator associated
with the stability of $\pfl(x)$. The homogeneous problem ${\cal L}\psi=0$
has the following two independent solutions,
\beq
\label{eq:psilin}
\psi_b (x) = \frac{1}{\cosh x}, \quad
\psi_u (x) = \frac{x}{\cosh x} + \sinh x,
\eeq
where $\psi_b(x) = \frac12 \frac{d}{dx} \pfl(x)$ is bounded and
$\psi_u(x)$ unbounded as $x \to \pm \infty$. By the variation-of-constants
method, we find the general solution to (\ref{eq:phi1}),
\begin{eqnarray}
\phi_1(x; A, B) &=&
\left[A + \frac12 \cosh x + \frac12 \int_0^x \frac{\xi}{\cosh \xi} d
  \xi\right]  \,\frac{1}{\cosh x} \nonumber\\
&&{}+[B - \arctan e^x]\left(\frac{x}{\cosh x} + \sinh x\right),\nonumber
\end{eqnarray}
with $A, B \in \mathbb{R}$. The solution $\phi_1(x)$ of (\ref{eq:phi1}) 
must be bounded as $x \to -\infty$ and is normalized by $\phi_1(0) = 0$
(since $\phi_h(0) = \pfl(0) = \pi$). Thus, we find that 
$A = -\frac12$ and $B = 0$.
Note that $\lim_{x \to - \infty} \phi_1(x) = 1$, 
which agrees with the fact that
$\lim_{x \to - \infty} \phi_h(x) = \arcsin \gamma = \gamma + \O(\gamma^3)$.
The solution $\phi_1(x)$ is clearly not bounded as $x \to \infty$,
the unbounded parts of $\phi_1(x)$ and $\frac{d}{dx} \phi_1(x)$ are
given by
\beq
\label{eq:phi1u}
\phi_1|_u (x) = - \arctan e^x \sinh x, \; \; \frac{d}{dx} \phi_1|_u
(x) = -\arctan e^x \cosh x.  \eeq It follows that $\phi_1(x) =
\O(\gamma^{-\sigma})$ for some $\sigma > 0$ if $e^x =
\O(\gamma^{-\sigma})$, i.e., if $x = \sigma |\ln \gamma|$ at leading
order. Using this, it is a straightforward procedure to show that the
rest term $\gamma^2 R_2(x; \gamma)$ in (\ref{eq:expphi}) is of
$\O(\gamma^{2-2\sigma})$ for $x = \sigma |\ln \gamma| + \O(1)$ (and
$\sigma > 0$). Hence, the approximation of $\phi_h(x)$ by expansion
(\ref{eq:expphi}) breaks down as $x$ becomes of the order $|\ln
\gamma|$. On the other hand, it also follows that $\phi_{\rm
  appr}^1(x) = \pfl(x) + \gamma \phi_1(x)$ is a uniform
$\O(\gamma)$-accurate approximation of $\phi_h(x)$ on an interval
$(-\infty, L]$ for $L = \frac12 |\ln \gamma| + \O(1)$. Since $\pfl(L)
+ \gamma \phi_1(L) = \O(\sqrt\gamma)$ for such
$L$, 
we can compute $L_{\pi} = \frac12 |\ln \gamma| + \O(1)$, as $L_\pi$ is
the value of $x$ at which
\[
0 = \frac{d}{dx} \phi_h (x) = \frac{d}{dx} \phi_{\rm appr}^1(x) + \O(\gamma) =
\frac{d}{dx} \pfl(x) + 
\gamma \frac{d}{dx} \phi_1|_u (x) + \O(\gamma).
\] 
We introduce $Y$ by $e^x = \frac{Y}{\sqrt{\gamma}}$, so that it follows
by (\ref{eq:funflux}) and (\ref{eq:phi1u}) that 
$Y = \frac{4}{\sqrt{\pi}} + \O(\sqrt{\gamma})$, i.e.
\[
L_\pi(\gamma) = \frac12 |\ln \gamma| + \ln \frac{4}{\sqrt{\pi}} + \O(\sqrt{\gamma}).
\]
A straightforward calculation shows that (in the new coordinates)
\[
\phi_h(L_\pi) = 2 \pi - 2 \sqrt{\pi} \sqrt{\gamma} + \O(\gamma).
\]
As $\phi_h(x)$ and $\pfl(x)$ both converge exponentially fast to 
fixed points which are order~$\gamma$ apart for $x\to-\infty$, it
follows immediately that $\phi_1(x;\gamma)= {\cal O}(1)$ and 
$R_2(x;\gamma) = {\cal O}(1)$, uniform for $x<0$.
\end{proof}

Now we are ready to consider the stability of the various types of
$\pi$-fluxons individually.

\subsection{Stability of the type~1 solution}

\begin{lemma}\label{lem:L1}
For all $0\leq\gamma<\gammacr$, all eigenvalues of ${\cal
    L}^1(x;\gamma)$ are strictly negative.  For $\gamma=\gammacr$, the
  operator ${\cal L}^1(x;\gammacr)$ has~0 as its largest eigenvalue.
  For $\gamma=0$, the largest eigenvalue is $-\frac1{4}(\sqrt 5 + 1)$.
  Furthermore, for all $0\leq\gamma<\gammacr$, the type~1
  semikinks~$\phi_\pi^1(x;\gamma)$ are Lyapunov stable in the
  following sense. For all $\eps>0$, there is some $\delta>0$ such
  that any solution $\phi(x,t)$ of the semi-fluxon
  equation~(\ref{eq:pSG}), which is convergent to $0$ at $x\to-\infty$
  and to $\pi$ at $x\to+\infty$ and which satisfies initially
  $\|\phi(\cdot,0)-\phi_\pi^1(\cdot;\gamma)\|_{H_1} +
  \|\phi_t(\cdot,0)\|_{L_2}<\delta$ will satisfy
  $\|\phi(\cdot,t)-\phi_\pi^1(\cdot;\gamma)\|_{L_2} +
  \|\phi_t(\cdot,t)\|_{L_2}<\eps$ for all $t\in\mathbb{R}$.
\end{lemma}
\begin{proof}
From Lemma~\ref{lem:zero} it follows that ${\cal L}^1$ has an eigenvalue $\Lambda=0$ at $\gamma=\gammacr$. The eigenfunction is $D_x\phi^1_\pi(x;\gammacr)$ and this function is always positive, since $\phi^1_\pi(x;\gammacr)$ is monotonically increasing. From Sturm's Theorem, it follows that $\Lambda=0$ is the largest eigenvalue of ${\cal L}^1$ at $\gamma=\gammacr$.
Next we consider $\gamma=0$. We can explicitly determine all
eigenvalues of ${\cal L}^1(x;0)$.  From the explicit expression for
$\phi_\pi^1$ it follows that ${\cal L}^1(x;0)$ is a continuous even
operator. For fixed $\Lambda$, the operator ${\cal L}^1(x;0)-\Lambda$
has two linearly independent solutions. Since the fixed point is a
saddle point and the decay rate to this fixed point is like $e^{-x}$,
there is one solution that is exponentially decaying at $+\infty$ and
there is one solution that is exponentially decaying at $-\infty$, if
$\Lambda>-1$. If we denote the exponentially decaying function at
$-\infty$ by $v_-(x;\Lambda)$, then the exponentially decaying
function at $+\infty$ up to a constant is given by
$v_+(x;\Lambda)=v_-(-x;\Lambda)$ (since ${\cal L}^1$ is symmetric in
$x$). Obviously, $v_+(0;\Lambda)=v_-(0;\Lambda)$, hence $\Lambda$ is
an eigenvalue if $D_xv_+(0;\Lambda)=D_xv_-(0;\Lambda)$, (i.e., when
$D_xv_-(0;\Lambda)=0$) or if $v_-(0;\Lambda)=0$.

Using \cite{mann97}, we can derive explicit expression
for the solutions~$v_-(x;\Lambda)$ (see also~\cite{derk03}). Using
$x_1=\ln(\sqrt2+1)$, we have
\[
v_-(x;0) = \sech(x-x_1),\quad
v_-(x;\Lambda) = e^{\mu (x-x_1)} \, [\tanh(x-x_1)-\mu] , \quad \mu =
\sqrt{\Lambda+1}.
\]
A straightforward calculation shows that $v_-(0;\Lambda)\neq0$. The
condition $D_xv_-(0;\Lambda)=0$ gives that
\[
\mu^2-\frac12\sqrt2 \mu -\frac 12 = 0 ,\qmbox{hence}
\sqrt{\Lambda+1}  = \frac14 \sqrt 2 (\sqrt 5 -1) \Rightarrow
\Lambda=-\frac14 (\sqrt 5 + 1).
\]

Now assume that the operator ${\cal L}^1(x;\gamma)$ has a positive
eigenvalue~$\Lambda^1(\gamma)$ for some $0\leq\gamma<\gammacr$. Since
$\Lambda$ depends continuously on $\gamma$, there has to be some
$0<\widehat\gamma<\gammacr$ such that
$\Lambda^1(\widehat\gamma)=0$. However, from Lemma~\ref{lem:zero} it
follows that this is not possible. 

Nonlinear or Lyapunov stability can be derived by
looking at the ``temporal Hamiltonian''
\[
{\cal H}(\phi,p) =
\int_{-\infty}^\infty \left[\frac12 p^2 + \frac 12 (\phi_x)^2
-\cos(\phi+\theta) - \gamma(\phi+\theta)\right]\, dx
\]
This functional is a Lyapunov function for the system~(\ref{eq:pSG}),
i.e., any solution $\phi(x,t)\in H^2(\mathbb{R})$ of~(\ref{eq:pSG})
satisfies $\frac{d}{dt} {\cal H}(\phi,\phi_t) =0$, hence ${\cal
  H}(\phi(\cdot,t),\phi_t(\cdot,t))={\cal
  H}(\phi(\cdot,0),\phi_t(\cdot,0))$ for any $t\in\mathbb R$.
Furthermore, the linearization $D^2{\cal H}$ at
$(\phi,p)=(\phi_\pi^1,0)$ (the point related to the $\pi$-fluxon) is
given by
\[
D^2{\cal H}(\phi_\pi^1,0) =
\left(
\begin{array}{cccc}
  -{\cal L}^1(x;\gamma) & 0 \\ 0& I
\end{array}
\right),
\]
which is a strictly positive definite self-adjoint operator on
$L_2(\mathbb R)\times L_2(\mathbb R)$ with domain $H_2(\mathbb
R)\times L_2(\mathbb R)$.  So there is some $c>0$ such that for any
$(\phi,p)\in H_2\times L_2$, we have $H(\phi,p)-H(\phi_\pi^1,0)\geq
c(\|\phi-\phi_\pi^1\|_{L_2}^2+ \|p\|_{L_2}^2)$, see e.g.~\cite{kato76,
  yosida95}. Finally, it is straightforward to prove that there is
some $C>0$ such that $H(\phi,p)-H(\phi_\pi^1,0)\leq
C(\|\phi-\phi_\pi^1\|_{H_1}^2+ \|p\|_{L_2}^2)$ for any $(\phi,p)\in
H_2\times L_2$.
\end{proof}


\begin{figure}[tbp!]
\begin{center}
\vspace{2cm}
\end{center}
\caption{(a) The eigenvalue of linear operator associated to the
  type~1 semifluxon as a function of 
  the bias current~$\gamma$. The dashed line is the boundary of the
  continuous spectrum. 
(b) A simulation of the evolution of a $\pi$-kink in the presence of a
bias current above the critical value ($\gamma>\gammacr$). The plot 
  presents the magnetic field $\phi_x$. The numerics show that the 
  instability leads to the release of wave trains of traveling wave
  fluxons. In this evolution a damping, which is proportional to 
  $\phi_t$, has been applied to the system.}   
\label{SFs_1}
\end{figure}

Using standard procedures in MATLAB, the eigenvalues of the type~1
$\pi$-fluxon have been calculated numerically as a function of the
applied bias current~$\gamma$ and are presented in
Figure~\ref{SFs_1}(a).  Further details of the computational procedure
are presented in section~\ref{3sec6}. Figure~\ref{SFs_1}(a) shows that
the type~1 semifluxon has only one eigenvalue.  This eigenvalue tends
to zero when the bias current~$\gamma$ approaches the critical
value~$\gammacr$ as has been derived analytically. 
It was first proposed in~\cite{kato97,kukl95} that a constant driving
force can excite the largest eigenvalue of a semifluxon toward zero.

When we apply a bias current above the critical value~$\gammacr$,
numerics show that the stationary $\pi$-kink bifurcates into a semifluxon that
reverses its polarity and releases a fluxon. This process
keeps repeating itself: the semifluxon changes its direction back and
forth with releasing a fluxon or antifluxon in every change. A simulation
of the release of fluxons from a semifluxon is presented in
Figure~\ref{SFs_1}(b). In experiments, the polarity of a semifluxon
can also be reversed by applying a magnetic field \cite{hilg03}.

When $\gamma=\gammacr$, the type~1 and type~3 semifluxons
coincide. From the numerical analysis of the eigenvalues of the type~3
semifluxon (see section~\ref{3sec33} for details), it follows that
there is an eigenvalue at the edge of the continuous spectrum for 
$\gamma=\gammacr$. We conjecture that this eigenvalue bifurcates into
the edge of the continuous spectrum at this point as $\gamma$ 
increases to $\gamma_{\rm cr}$ (see Figure \ref{SFs_3}).
 
\subsection{Instability of type~2 solutions}

\begin{lemma}\label{lem:L2}
  For all $0<\gamma<\gamma^*$, the largest
  eigenvalue 
  of ${\cal L}^2(x;\gamma)$ is strictly positive. In the limit
  $\gamma\to0$, the largest eigenvalue 
  of ${\cal L}^2(x;\gamma)$ converges to $\frac14(\sqrt5-1)$.
\end{lemma}

\begin{proof}
  Using the approximation for the homoclinic orbit $\phi_h(x;\gamma)$
  in Lemma~\ref{lem:hom}, we see that, for $\gamma$ small, an
  approximation for the $\pi$-fluxon of type~2 is given by (as before,
  $x_1=\ln(1+\sqrt2)$)
\begin{equation}
\label{eq:semifluxon2}
\hspace{0.7cm}\phi_\pi^2(x;\gamma) 
=\left\{
\begin{array}{ll}
\pfl(x+x_1) + {\cal O}(\gamma),&x<0\\
\pi+\pfl(\widetilde x)+\gamma\phi_1(\widetilde x) + \gamma^2 R_2(\widetilde x;\gamma), 
&0<x<L_\pi(\gamma)+x_1\\   
\pi+\pfl(-\widehat x)+\gamma\phi_1(-\widehat x) + \gamma^2
R_2(-\widehat x;\gamma),&x>L_\pi(\gamma)+x_1
\end{array}\right.
\end{equation}
with $\widetilde x=x-x_1$ and $\widehat x=x-2L_\pi(\gamma)-x_1$.

There is no limit for $\gamma\to0$, since the semifluxon breaks in two
parts, one of them being the $3\pi$-fluxon~$\phi_{3\pi}^2(x;0)$. In a
similar way as we found the largest eigenvalue for the linearization
operator ${\cal L}^1(x;0)$ about the $\pi$-fluxon~$\phi_\pi^1(x;0)$,
we can find the largest eigenvalue for the linearization operator
${\cal L}^2(x;0)$ about the $3\pi$-fluxon~$\phi_{3\pi}^2(x;0)$. The
largest eigenvalue is $\Lambda^2(0)=\frac14(\sqrt5-1)$ and the
eigenfunction is
\[
\psi^2(x;0) = \left\{
  \begin{array}{lll}
  e^{\mu_0(x+x_1)}(\mu_0-\tanh(x+x_1)),&& x<0,\\
  e^{\mu_0(-x+x_1)}(\mu_0-\tanh(-x+x_1)),&& x>0,
  \end{array}
\right.
\]
where $\mu_0=\sqrt{\Lambda^2(0)+1}=\frac14\sqrt2(1+\sqrt5)$. (It can
be shown that there is another smaller eigenvalue $\Lambda=-\frac12$
and similar eigenfunction if $\mu=\frac12\sqrt2=\tanh(x_1)$, see
Remark~\ref{rem:3.8}.) 

In a similar way, using the approximation~(\ref{eq:semifluxon2}) for
$\phi_\pi^2(x;\gamma)$, the eigenfunction of an eigenvalue of
$\phi_\pi^2$ for $\gamma$ small is approximated by
\[
\psi^2(x;\gamma) = \left\{
  \begin{array}{lll}
    e^{\mu(x+x_1)}(\mu-\tanh(x+x_1))+ {\cal O}(\sqrt\gamma),&&x<0\\
   k_2\,e^{-\mu\widetilde x}(\mu-\tanh(-\widetilde x))+
   k_3\,e^{\mu\widetilde x}(\mu-\tanh\widetilde x)+ {\cal
     O}(\sqrt\gamma),
    &&0<x<L_\pi(\gamma)+x_1\\
    k_4\,e^{\mu(-\widehat x)}(\mu+\tanh\widehat x) + {\cal
      O}(\sqrt\gamma), && x>L_\pi(\gamma)+x_1.
  \end{array}\right.,
\]
where $k_i$ and $\mu$ have to be determined. The eigenvalue $\Lambda$
follows from $\mu=\sqrt{\Lambda^2+1}$.  Note that the secular term which
is growing at infinity with the multiplication factor~$k_3$ is
included in this approximation. When $\gamma=0$ and $k_3=0$, the first
two lines in the definition of $\psi^2$ are the eigenfunction of the
linearized problem about the heteroclinic connection between $0$ and
$3\pi$, as presented above. 
When $\gamma$ is nonzero, $k_3$ can be of order ${\cal
   O}(\gamma^\sigma)$ for $\sigma>\frac\mu2$ as the secular term is
 of order ${\cal O}(\gamma^{-\mu/2})$ at $x=L_\pi(\gamma)+x_1$.

The constants $k_2,~k_3$ and $k_4$ and the parameter~$\mu$ have to be
chosen such that for $\gamma>0$ (but small) the
function~$\psi^2(x,\gamma)$ is continuously differentiable at $x=0$ and
$x=L_\pi(\gamma)+x_1$. From the continuity conditions at $x=0$, we obtain:
\begin{eqnarray*}
  k_2&=&\frac{\sqrt2}{4\mu(\mu-1)(\mu+1)}+{\cal O}(\sqrt\gamma),\\
  k_3 &=&
  \frac{(3+2\sqrt2)^\mu(2\mu^2-\mu\sqrt2-1)(2\mu-\sqrt2)}%
  {4\mu(\mu^2-1)}  +  {\cal O}(\sqrt\gamma).
\end{eqnarray*}

From one of the continuity conditions at $x=L_\pi(\gamma)+x_1$, we
obtain $k_4=k_4(k_2,k_3,\mu)$. Now we are left with one more matching
condition. Values of $\mu$ for which this condition is satisfied
correspond to the eigenvalues of the operator ${\cal L}^2(x;\gamma)$
for $\gamma$ small. More explicitly, the spectral parameter $\mu$ has
to satisfy the equation
\begin{eqnarray} {\cal F}(\mu)= 16^\mu
  k_3(\mu-1)^2(\gamma\pi)^{-\mu}((3\mu+4)\pi \gamma+16\mu)+{\cal
    O}(\gamma^{-\mu+2})=0.
\end{eqnarray}
Note that this expression is not defined at $\gamma=0$. This
corresponds to the singularities in the expression for $\phi^2$ as
$\gamma\to 0$ due to the fact that
$L_\pi(\gamma)\to\infty$ for $\gamma\to 0$.  Evaluating ${\cal
  F}(\mu)\,(\gamma\pi)^{\mu}$ at $\gamma=0$, we see that there are
four positive roots for $\mu$, leading to four squared eigenvalues,
namely $\Lambda(0)=\frac14(\sqrt5-1)$, $-\frac12$, and the double
eigenvalue $\Lambda(0)=0$. The first two come from the zeros of $k_3$
and are related to the eigenvalues of the $3\pi$-fluxon. The double
zero eigenvalues are the eigenvalues of the fluxon. One can also
notice that there is no term with a multiplication factor $k_2$ to
this leading order. This term appears at most of order ${\cal
  O}(\gamma^{\mu+2})$. Finally, as with the type~1 semi-fluxon, the root
$\mu=0$ corresponds to the edge of the continuous spectrum and the
``eigenfunction'' is not in $H_2(\mathbb{R})$.

The proof that the largest eigenvalue is near $\frac14(\sqrt5-1)$ for
$\gamma$ small will be complete if we can show that ${\cal
  F}_\mu(\sqrt2/4(1+\sqrt5))\neq0$, i.e. the non-degeneracy condition
that says that the eigenvalue can be continued continuously for
$\gamma$ small.

Simple algebraic calculations give that
\begin{equation} {\cal F}_\mu(\frac{\sqrt2}{4}(1+\sqrt5)) =
  c_1\gamma^{-\frac{\sqrt2}{4}(1+\sqrt5)}+{\cal
    O}(\gamma^{1-\frac{\sqrt2}{4}(1+\sqrt5)})
\end{equation}
with $c_1$ a positive constant. Hence, ${\cal
  F}_\mu(\frac{\sqrt2}{4}(1+\sqrt5))>0$.

This completes the proof that the largest eigenvalue is near
$\frac14(\sqrt5-1)$ for small but positive~$\gamma$. Since the largest
eigenvalue depends continuously on $\gamma$, it can only disappear at
a bifurcation point. There are no bifurcation points and it is not
possible that the eigenvalue becomes 0 (see Lemma~\ref{lem:zero}),
hence the largest eigenvalue will be positive as long as fluxon
$\phi^2_\pi(x;\gamma)$ exists, i.e., for $0<\gamma<\gamma^*$.
\end{proof}

\begin{remark} \rm
We cannot use a comparison theorem, because $\phi_\pi^2<\phi_\pi^3$
for $x<0$ and $\phi_\pi^2>\phi_\pi^3$ for $x>0$.
\end{remark}

\begin{figure}[tb!]
\begin{center}
\vspace{2cm}
\end{center}
\caption{(a) The eigenvalues of the linear operator associated to the
  type 2~semifluxon as a function of the bias current~$\gamma$. The
  result that the largest eigenvalue is always positive shows the
  instability of type 2 semifluxon. When $\gamma\to0$,
  $\Lambda\to\frac14(\sqrt{5}-1)$ which is the largest eigenvalue of a
  $3\pi$-kink. At $\gamma=0$, one eigenvalues comes out of the edge of
  the continuous spectrum (dashed line). 
(b) The evolution of a $3\pi$-kink (\ref{sge2b}) for
$\gamma=0$. The plot is presented in terms of the magnetic field
$\phi_x$.The separation of a fluxon from the semifluxon can be clearly
seen.}
\label{SFs_2}
\end{figure}

To consider the relation between the eigenvalues of ${\cal
  L}^2(x;\gamma)$ and the stability problem of $\phi^2_\pi(x;\gamma)$,
we denote the largest eigenvalue of ${\cal L}^2(x;\gamma)$ by
$\Lambda^2(\gamma)$. The associated eigenvalues for the linearizations are
solution of the equation $ \lambda^2-\Lambda^2(\gamma)=0$, hence
$\lambda = \pm \sqrt{\Lambda^2(\gamma)}.  $ Since
$\Lambda^2(\gamma)>0$, this implies that one of the two eigenvalues
has positive real part, hence the $\pi$-fluxons of type~2 are unstable. The
numerically obtained eigenvalues of semifluxons of this type as a
function of $\gamma$ are shown in Figure~\ref{SFs_2}(a). 
In the proof of Lemma~\ref{lem:L2} we have found three different
eigenvalues for $\gamma$ small and the possibility of a fourth
eigenvalue coming out of the continuous spectrum at $\gamma=0$. In
Figure~\ref{SFs_2}(a), we see the  continuation of those
eigenvalues. 
In Figure~\ref{SFs_2}(b), we present the evolution of a $3\pi$-kink
(\ref{sge2b}) which is the limit of a type 2 semifluxon when
$\gamma\to0$. The separation of a fluxon from the semifluxon is
clearly seen and indicates the instability of the state 
(which confirms the analysis in the proof of Lemma \ref{lem:L2}).

\begin{remark} \label{rem:3.8}
\rm A type 2 semifluxon can be seen as a concatenation of
  a $3\pi$- and a $-2\pi$-kink in the limit
  $\gamma\to0$. In that limit the other eigenvalues of
  ${\cal L}^2(x;\gamma)$ converge to 0, $-\frac12$, and $-1$. The
  eigenvalues 0 and $-1$ are contributions of the $-2\pi$-kink. The
  eigenvalue $-\frac12$ corresponds to the first excited
  state of the $3\pi$-kink with eigenfunction
\[
\psi^2(x;0) = \left\{
  \begin{array}{lll}
  e^{\mu(x+x_1)}(\mu-\tanh(x+x_1)),&& x<0,\\
  e^{\mu(-x+x_1)}(\tanh(-x+x_1)-\mu),&& x>0,
  \end{array}
\right.
\]
where $\mu=\sqrt{\Lambda+1}=\frac{1}{\sqrt2}$. 
\end{remark}

\subsection{Instability of type~3 solutions}
\label{3sec33}

\begin{lemma}\label{lem:L3}
For all $0<\gamma<\gammacr$, the largest eigenvalue 
of  ${\cal L}^3(x;\gamma)$ is strictly positive. 
For $\gamma=\gammacr$, the operator ${\cal L}^3(x;\gammacr)$ has 
0 as its largest eigenvalue. 
\end{lemma}
\begin{proof}
The solution~$\phi^3_\pi(x,\gammacr)=\phi^1_\pi(x,\gammacr)$, hence from
Lemma~\ref{lem:L1} it follows that the largest eigenvalue is
$\Lambda=0$. 

For $\gamma$ near zero, we will use the approximation for the
homoclinic orbit $\phi_h(x;\gamma)$ in Lemma~\ref{lem:hom} to get an
approximation for the type~3 fluxon
\[
\phi_\pi^3(x;\gamma) 
=\left\{\renewcommand{\arraystretch}{1.3}
\begin{array}{ll}
\phi_{\rm appr}^1(\widehat x) = 
\pfl(\widehat x)+\gamma\phi_1(\widehat x) + \gamma^2 
R_2(\widehat x;\gamma), & x<-L_\pi(\gamma)+x_1,\\
\phi_{\rm appr}^2(\widetilde x) = 
\pfl(-\widetilde x)+\gamma\phi_1(-\widetilde x) + \gamma^2
R_2(-\widetilde x;\gamma), &-L_\pi(\gamma)+x_1<x<0,\\   
\phi_{\rm appr}^3(-x-x_1)=\pi+\pfl(-x-x_1) + {\cal O}(\gamma), &x>0,
\end{array}\right.
\] 
where $\widetilde x= x-x_1$ and $\widehat x=x-x_1+2L_\pi(\gamma)$.

In the limit $\gamma\to 0$, the type~3 semi-fluxon break into a type~1
semi-fluxon and a fluxon. Both are stable and the largest eigenvalue
of the fluxon is zero, while the largest eigenvalue of the type~1
semi-fluxon is negative. Hence to approximate the  largest eigenvalue
of the type~3 semi-fluxon for $\gamma$ small, we set 
\[
\Lambda(\gamma) = \gamma \La_1(\gamma).
\]
To construct the first part of the approximation of the eigenfunction, we
consider $x<-L_\pi(\gamma)+x_1$, i.e., $\widehat x<L_\pi(\gamma)$. In this
part of the arguments, we will drop the hat in $\widehat x$.
On $(-\infty, L_\pi)$, 
we expand $\psi_{\rm approx}^1 
= \psi_0 + \gamma \psi_1$
, this yields the following equations for $\psi_{0,1}(x)$,
\beq
\label{eqspsi01}
\L \psi_0 =  0, \; \;
\L \psi_1 = [\La_1(0) - \phi_1(x) \sin \pfl(x)] \psi_0.
\eeq
As $\psi_{\rm approx}^1$ has to be an eigenfunction, we have $\psi_{\rm approx}^1
(x) \to 0$ as $x \to -\infty$. Furthermore, we remove the scaling
invariance by assuming that $\psi_{\rm approx}^1 (0) = 1$.
This implies that $\psi_0(x)$ is given by
\beq\label{solpsi0}
\psi_0(x) = \frac{1}{\cosh x} 
\eeq 
(see \ref{eq:psilin}). To solve the
$\psi_1$-equation, we note that $\frac{d}{dx} \phi_1(x)$ is a solution
of
\[
\L \psi = - \phi_1 \sin \pfl \frac{d}{dx} \pfl = 
- 2 \phi_1 \sin \pfl \psi_0,
\]
(see (\ref{eq:phi1}) and (\ref{eq:funflux})) so that we find as
general solution, \renewcommand{\arraystretch}{1.3}
\begin{eqnarray}
\psi_1(x) &=&
[A - \frac12 \La_1 (\ln(\cosh x) + 
\int_0^x \frac{\xi}{\cosh^2 \xi} d \xi)] 
\frac{1}{\cosh x}\nonumber\\
&&{}+\left[B + \frac12 \La_1 \tanh x\right]
\left(\frac{x}{\cosh x} + \sinh x\right)
+ \frac12 \frac{d}{dx} \phi_1.\nonumber
\end{eqnarray}
Using $\lim_{x \to -\infty} \psi_1(x) = 0$ and $\psi_1(0) = 0$
we find that $A = \frac{\pi}{4}$, $B = \frac12 \La_1(0)$.
As in the case of $\phi_1(x)$, we are especially interested in the
unbounded parts of $\psi_1(x)$ and $\frac{d}{dx} \psi_1(x)$, 
\beq
\label{eq:psi1u}
\begin{array}{rcl}
\psi_1|_u (x) & = & \frac12 \La_1 (1 + \tanh x) \sinh x
- \frac12 \arctan e^x \cosh x, 
\\
\frac{d}{dx} \psi_1|_u (x) & =  & \frac12 \La_1 (1 + \tanh x) \cosh x
- \frac12 \arctan e^x \sinh x.
\end{array}
\eeq  
We note that the error term 
$|\psi(x) - \psi_{\rm appr}^1(x)| = \gamma^2|S_2(x;\gamma)|$ is at most $\O(\gamma)$
on $(-\infty, L_\pi)$ (the analysis is similar to that for 
$\gamma^2|R_2(x;\gamma)|$).

Next consider the second part of the approximation, i.e., $x$ between
$-L_\pi(\gamma)+x_1$ and 0. Here we define the translated coordinate
$\widetilde x = x-x_1$, which is on the interval $(-L_{\pi}, -x_1)$
and again we drop the tildes.  Since we have to match $\psi_{\rm
  appr}^1(x)$ to the approximation $\psi_{\rm appr}^2(x)$ of
$\psi(x)$, along $\phi_{\rm appr}^2(x)$ and thus defined on the
interval $(-L_{\pi},-x_1)$, we need to compute $\psi_{\rm
  appr}^1(L_\pi)$ and $\frac{d}{dx} \psi_{\rm appr}^1(L_\pi)$ which to
the leading order are calculated from (\ref{eq:psi1u}), i.e.\ 
\beq
\label{psiLpi}\hspace{1cm}
\psi_{\rm appr}^1(L_\pi) = \frac{2 \La_1(0)}{\sqrt{\pi}} \sqgam +
\O(\gamma), \quad \frac{d}{dx} \psi_{\rm appr}^1(L_\pi) = \frac{2
  \La_1(0) - \pi}{\sqrt{\pi}} \sqgam + \O(\gamma).  
\eeq 
Thus, both
$\psi_{\rm appr}^1(L_\pi)$ and $\frac{d}{dx} \psi_{\rm appr}^1(L_\pi)$
are $\O(\sqgam)$.  

Now, we choose a special form for $\psi_{\rm
  appr}^2(x)$, the continuation of $\psi(x)$, i.e.  the part
linearized along $\phi_{\rm appr}^2(x)$. It is our aim to determine
the value of $\La_1$, for which there exists a positive integrable
$C^1$ solution $\psi$ of $\L^3(x;\gamma)\psi=\gamma\La_1(0)\psi$. By
general Sturm-Liouville theory \cite{Titch} we know that this value of
$\La_1$ must be the largest eigenvalue.
Our strategy is to try to continue $\psi(x)$ beyond $(-\infty, L_\pi)$
by a function that remains at most $\O(\sqgam)$, i.e. we do not
follow the approach of the existence analysis and thus do not
reflect and translate $\psi_{\rm appr}^1(x)$ to construct 
$\psi_{\rm appr}^2(x)$ (since this solution becomes in general 
$\O(1)$ for
$x = \O(1))$. Instead, we scale $\psi_{\rm appr}^2(x)$ as $\gamma \tpsi(x)$.
The linearization $\tpsi(x)$ along 
$\phi_{\rm appr}^2(x)$ on the interval 
$(-L_{\pi}, x_1)$ must solve 
$\L \tpsi = O(\gamma)$, thus, at leading order
\beq
\tpsi(x) = \frac{\tilde{A}}{\cosh x} + \tilde{B}\left(\frac{x}{\cosh
    x} + \sinh x\right). 
\eeq
The approximation $\psi_{\rm appr}^2(x) = \gamma \tpsi(x)$ must be matched to 
$\psi_{\rm appr}^1(L_\pi)$ and $\frac{d}{dx} \psi_{\rm appr}^1(L_\pi)$
at $x = -L_\pi$, i.e.
\[ 
\frac{2 \La_1(0)}{\sqrt{\pi}} = - \frac{2 \tilde{B}}{\sqrt{\pi}} + \O(\sqgam), 
\; \; 
\frac{2 \La_1(0) - \pi}{\sqrt{\pi}} = \frac{2 \tilde{B}}{\sqrt{\pi}} 
+ \O(\sqgam).
\]
Note that $\tilde{A}$ does not appear in these equations; as a
consequence, $\psi_{\rm appr}^1(x)$ and $\psi_{\rm appr}^2(x)$ can
only be matched for a special value of $\La_1$, $\La_1(0) = \frac14
\pi$, with $\tilde{B} = -\La_1(0) < 0$.  Thus for this special value
of $\La_1$ and for $\tilde{A} > 0$, we have found a positive
$C^1$-continuation of the solution $\psi(x)$ of the eigenvalue problem
for $\L^3(x;\gamma)$ -- recall that $x < 0$ in the domain of
$\tpsi(x)$. At the point of discontinuity ($-x_1$ for $\tpsi(x)$, or
at $x = 0$ in the original coordinates of (\ref{eq:pSG})), we have
\beq
\label{tpsidisc}\renewcommand{\arraystretch}{1.5}
\begin{array}{rcrcr}
&&\psi_{\rm appr}^2(-x_1)  =  \gamma \tpsi(-x_1)  =  
\gamma[\frac12 \sqrt{2} \tilde{A} - \frac{\pi}{8} \sqrt{2}
(\ln(\sqrt{2} - 1) - \sqrt{2})] + \O(\gamma^2), 
\\
&&\frac{d}{dx} \psi_{\rm appr}^2(-x_1)  =  
\gamma \frac{d}{dx} \tpsi(-x_1)  = 
\gamma[\frac12 \tilde{A} - \frac{\pi}{8} 
(\ln(\sqrt{2} - 1) + 3\sqrt{2})] + \O(\gamma^2).
\end{array}
\eeq
Hence, we have constructed for a special choice of $\La$,
$\La = \La_{\ast}= \frac\pi4 \gamma + \O(\gamma\sqrt\gamma) > 0$,
an approximation of a family of positive solutions of
the eigenvalue problem for $\L^3(x;\gamma)$ on $x < 0$ -- 
in the coordinates of (\ref{eq:pSG}) -- that attain the values
given by (\ref{tpsidisc}) at $x = 0$, and that 
decay to $0$ as $x \to -\infty$. The question is now whether we
can `glue'  
an element of this family in a $C^1$-fashion
to a solution of the eigenvalue problem for $\L^3(x;\gamma)$ 
on $x > 0$ -- with $\La = \La_{\ast}$ --
that decays (exponentially)
as $x \to \infty$. If that is possible, we have 
constructed a positive integrable solution to the eigenvalue problem for
$\L^3(x;\gamma)$, which implies that
$\La_{\ast} > 0$ is the critical eigenvalue and that $\pp3(x)$
is unstable.

An approximation of $\psi(x)$ on $x > 0$, $\psi_{\rm appr}^3(x)$,
is obtained by linearizing along $\phi_{\rm appr}^3(x)$ and by
translating $x$ so that $x \in (x_1, \infty)$.
Since $\psi_{\rm appr}^3(x)$ has to match to expressions of $\O(\gamma)$
(\ref{tpsidisc}) at $x_1$, 
we also scale $\psi_{\rm appr}^3(x)$,
$\psi_{\rm appr}^3(x) = \gamma \hat{\psi}(x)$. We find
that $\L \hat{\psi} = \O(\gamma)$ so that 
$\hat{\psi}(x)$ again has to be (at leading order) a linear combination 
of $\psi_b(x)$ and $\psi_u(x)$ (\ref{eq:psilin}). However,
$\hat{\psi}$ must be bounded as $x \to \infty$, which yields
that $\hat{\psi}(x) = \hat{A}/\cosh x + \O(\gamma)$ for some 
$\hat{A} \in \mathbb{R}$. At the point of discontinuity we thus have
\beq
\label{hpsidisc}
\begin{array}{rcrcr}
\psi_{\rm appr}^3(x_1) & = & \gamma \hat{\psi}(x_1) & = & 
\frac12 \sqrt{2} \hat{A} \gamma + \O(\gamma^2), \\
\frac{d}{dx} \psi_{\rm appr}^3(x_1) 
& = & \gamma \frac{d}{dx} \tpsi(x_1) & = &
-\frac12 \hat{A} \gamma + \O(\gamma^2).
\end{array}
\eeq
A positive $C^1$-solution of the eigenvalue problem for $\L^3(x;\gamma)$
exists  (for $\La = \La_{\ast}$) if there exist $\tilde{A}, \hat{A} > 0$
such that (see~(\ref{tpsidisc}) and~(\ref{hpsidisc}))
\beq
\label{eqtAhA}
\begin{array}{rcrcr}
\frac12 \sqrt{2} \tilde{A} & - & \frac{\pi}{8} \sqrt{2}
(\ln(\sqrt{2} - 1) - \sqrt{2}) & = &
\frac12 \sqrt{2} \hat{A} \\
\frac12 \tilde{A} & - & \frac{\pi}{8}
(\ln(\sqrt{2} - 1) + 3\sqrt{2}) & = &
-\frac12 \hat{A}
\end{array}
\eeq    
Since the solution of this system is  
given by $\tilde{A} = \frac14 \pi [\sqrt{2} + \ln(\sqrt{2} - 1)] > 0$ and 
$\hat{A} = \frac12 \pi \sqrt{2} > 0$, we conclude
that the eigenvalue problem for the $\pi$-fluxon $\pp3(x; \ga)$ 
has a positive largest eigenvalue 
\begin{equation}
\La_\ast = \frac\pi4\gamma + \O(\gamma\sqgam).
\label{EV_SFs_3}
\end{equation}

Hence the eigenvalue for $\gamma$ small is positive.
From Lemma~\ref{lem:zero} it follows that there are no zero eigenvalues
between 0 and $\gammacr$, hence 
the largest eigenvalue of $\L^3(\gamma)$ is positive for all values of
$\gamma$. 
\end{proof}

\begin{figure}[tb!]
\begin{center}
\vspace{2cm}
\end{center}
\caption{The eigenvalues of the the linear operator associated to the
  type 3 semifluxon as a function of the bias current~$\gamma$. The
  result that the largest eigenvalue is 
  always positive shows the instability of type 3 semifluxon. When
  $\gamma\ll1$, according to~(\ref{EV_SFs_3}) the largest
  eigenvalue is approximated by $\Lambda=\frac\pi4\gamma$ shown in
  dash-dotted line. The dashed line is the boundary of the continuous
  spectrum. 
}
\label{SFs_3}
\end{figure}

\begin{remark} \rm For any $\la = \O(\sqgam)$, or equivalently any $\La_1
  = \O(1)$, there exists a (normalized) solution to the eigenvalue
  problem for $\L^3(x;\gamma)$ on $x < 0$ that decays as $x \to
  -\infty$, and that is approximated by $\psi_{\rm appr}^1(x)$ and
  $\psi_{\rm appr}^2(x)$ (matched in a $C^1$-fashion at $\pm L_\pi$).
  If $\La_1$ is not $\O(\sqgam)$ close to $\frac14 \pi$, however,
  $\psi_{\rm appr}^2(x)$ cannot be scaled as $\gamma \tpsi(x)$ and the
  solution is not $\O(\gamma)$ at the point of discontinuity -- in
  general it is $\O(1)$. Moreover, for any $\La_1 = \O(1)$, there also
  exists on $x > 0$ a $1$-parameter family of (non-normalized)
  eigenfunctions for the eigenvalue problem for $\L^3(x;\gamma)$ that
  decay as $x \to \infty$. In this family there is one unique solution
  that connects continuously to the (normalized) solution at $x < 0$.
  In fact, one could define the jump in the derivative at $x=0$,
  ${\cal J}(\la; \gamma)$, as an Evans function expression (note that
  ${\cal J}(\la; \gamma)$ can be computed explicitly at $\gamma=0$,
  see~\cite{derk03}).  By definition, $\la^2$ is an eigenfunction of
  $\L^3(x;\gamma)$ if and only if ${\cal J}(\la; \gamma) = 0$.  In the
  above analysis we have shown that ${\cal J}(\la_{\ast}; \gamma) = 0$
  for $\la_{\ast}=\frac12\sqrt{\pi\gamma} + {\cal O}(\gamma)$.
\end{remark}

\begin{remark} \rm
The classical, driven, sine-Gordon equation, 
i.e. $\th \equiv 0$ and $\gamma\neq0$ in (\ref{eq:pSG}),
has a standing pulse solution, that can be seen, especially for 
$0 < \gamma \ll 1$, as a fluxon/anti-fluxon pair. This solution is  
approximated for $\frac{d}{dx} \phi > 0$ (the fluxon) by
$\phi_{\rm appr}^1(x)$ and for 
$\frac{d}{dx} \phi < 0$ (the anti-fluxon) by
$\phi_{\rm appr}^1(-x)$. It is (of course) unstable,
the (approximation of the) critical unstable eigenvalue can be obtained
from (\ref{psiLpi}). The corresponding
eigenfunction is approximated by
$\psi_{\rm appr}^1(x)$ on $(-\infty, L_\pi)$, 
and we conclude from (\ref{psiLpi}) that 
$\frac{d}{dx} \psi_{\rm appr}^1(L_\pi) = 0$ 
for $\la^2 = \gamma \La_1 = \gamma \frac{\pi}{2} + \O(\gamma \sqgam)$ (while
$\psi_{\rm appr}^1(L_\pi) > 0$). 
Hence, for this value of $\La_1$, we can
match $\psi_{\rm appr}^1(x)$ to 
$\psi_{\rm appr}^2(x) = \psi_{\rm appr}^1(-x)$ in a $C^1$-fashion,
it gives a 
uniform $\O(\gamma)$-approximation of the critical, 
positive (even, `two-hump') eigenfunction  
of the fluxon/anti-fluxon pair at the eigenvalue
$\la_+ = \frac12 \sqrt{2\pi} \sqgam + \O(\gamma) > 0$.
\end{remark}

To consider the relation between the eigenvalues of ${\cal
  L}^3(x;\gamma)$ and the stability problem of $\phi^3_\pi(x;\gamma)$,
we denote the largest eigenvalue of ${\cal L}^3(x;\gamma)$ by
$\Lambda^3(\gamma)$. The associated eigenvalues for the linearizations
are solution of the equation $ \lambda^2
-\Lambda^3(\gamma)=0,$ hence $\lambda = \pm \sqrt{\Lambda^3(\gamma)}.
$
Since 
$\Lambda^3(\gamma)>0$, this implies that
one of the two eigenvalues has positive real part, hence the fluxons of type~3
are unstable. In Fig.~\ref{SFs_3}, we present numerical calculations of the 
eigenvalues of type 3 semifluxon as a function of the bias current $\gamma$.

\begin{remark} \rm A type 3 semifluxon can be seen as a concatenation of
  a $2\pi$- and a $-\pi$-kink in the limit
  $\gamma\to0$. In that limit the other eigenvalue of
  ${\cal L}^3(x;\gamma)$ converges to $-\frac14(\sqrt5+1)$
  (Figure \ref{SFs_3}) which is a  contribution of the $-\pi$-kink.
\end{remark}

\section{Lattice $\pi$-kinks and their spectra in the  
continuum limit}
\label{3sec4}

In this section, we consider~(\ref{cont_appr}) for a small lattice
spacing~$a$, i.e., the driven $0$-$\pi$ sine-Gordon equation with a
small perturbation due to lattice spacing effects.
For $a=0$, the semifluxons of all types are constructed as
heteroclinic connections with transversal intersections at $x=0$ in
the two-dimensional phase space of the static
equation~(\ref{cont_appr_static}). Therefore, all three types of
semifluxons will still exist in the perturbed system with $0<a\ll1$,
see~\cite{guck86}.  The three types of semifluxons are denoted as
$\phi^i_\pi(x;a;\gamma)$, for $i=1,\,2$ and $3$. 
In Fig.~\ref{pplot}, we present the phase
portraits of the sine-Gordon equation both with and without the effect
of a perturbation due to lattice spacing.

\begin{figure}[tb!]
\begin{center}
\vspace{2cm}
\end{center}
\caption{The phase portrait of the stationary
  system~(\ref{cont_appr_static}) for $\gamma=0$ and some values of
  the lattice spacing~$a$. The dashed lines are the unperturbed phase
  portrait for $a=0$ and the other lines correspond to $a=0.5$.}
\label{pplot}
\end{figure}

The lattice spacing~$a$ does not affect the stationary points of the
phase portraits, as can be easily checked.  
The existence parameters~$\gamma^*$ and $\gammacr$ will be influenced
by the lattice spacing~$a$. For $a$ small, they are 
\begin{eqnarray}
\gamma^*(a) &=& \frac2{\sqrt{4+\pi^2}} + 
\underbrace{\frac{2\pi}{3(\pi^2+4)^2}}_{\approx0.0109}a^2 + 
{\cal O}(a^4),\label{gs}\\
\gammacr(a) &=& \frac2\pi + 
\underbrace{\frac{\sqrt{\pi^2-{4}} - \pi + 
2\arcsin(\frac{2}{\pi})}{3\pi^2}}_{\approx0.0223}a^2 + {\cal O}(a^4),
\label{ga}
\end{eqnarray}
see~\cite{susa04} for details.
For $\gamma>\gammacr(a)$ no static semifluxon exists.

As we have seen in the last section, for $a=0$, the type~3 semifluxon
is marginally unstable at $\gamma=\gammacr$ and $\gamma$ near zero. So
there is a possibility that lattice spacing effects stabilize the
type~3 semifluxon near those values of $\gamma$. However, it turns out
that this is not the case and the stability of the semifluxons is
similar to the case~$a=0$.
\begin{theorem}
  \label{th:stab2}
For $a$ small, the linearizations about the $\pi$-kinks have the following
properties:
\begin{romannum}
\item The eigenvalues of the linearization about the monotonic type~1
  $\pi$-kink $\phi_\pi^1(x;a;\gamma)$ are strictly negative for
  $0\leq\gamma<\gammacr(a)$. At $\gamma=\gammacr(a)$, the largest
  eigenvalue is zero. These $\pi$-kinks are linearly stable.
\item The largest eigenvalue of the linearization about the monotonic
  type~2 $\pi$-kink $\phi_\pi^2(x;a;\gamma)$ is strictly positive for
  $0<\gamma<\gamma^*(a)$. These $\pi$-kinks are linearly unstable.
\item The largest eigenvalue of the linearization about the monotonic
  type~3 $\pi$-kink $\phi_\pi^3(x;a;\gamma)$ is strictly positive for
  $0<\gamma<\gammacr(a)$. These $\pi$-kinks are linearly unstable. In the
  limit for $\gamma\to0$ and $\gamma\to\gammacr(a)$, the largest
  eigenvalue converges to zero.
\end{romannum}
\end{theorem}
The proof of this theorem will proceed along similar lines as the
proof in the previous section. 
First we consider the eigenvalue problem of a solution
$\phi^i_\pi(x;a;\gamma)$, which can be written as
\[
{\cal L}^i(x;a;\gamma) \, v = \lambda^2\, v ,
\]
where ${\cal L}^i(x;a;\gamma)$ is now defined as the linearization
associated with~(\ref{cont_appr_static}), i.e., 
\[
\begin{array}{lll}
{\cal L}^i(x;a;\gamma) &=&
D_{xx}-\cos\left(\phi^i_\pi(x;a;\gamma)+\theta(x)\right) 
\\
&&{}- \frac{a^2}{12}\left[
2\cos\widetilde\phi \,D_{xx} -
2(\phi^i_\pi(x;0;\gamma))_x\sin\widetilde\phi\,D_x \right.\\
&&\left.- 1 + 2\gamma\sin\widetilde\phi -
  ((\phi^i_\pi(x;0;\gamma))_x)^2\cos\widetilde\phi \right] + {\cal O}(a^4)
\end{array}
\]
where $\widetilde\phi=\phi^i_\pi(x;0;\gamma)+\theta(x)$.

Lemma~\ref{lem:zero} can be extended to $a\neq0$ and give a
necessary and sufficient condition for ${\cal L}^i(x;a;\gamma)$ to
have an eigenvalue $\Lambda=0$.
\begin{lemma}\label{lem:zero_a}
The eigenvalue problem 
\[
{\cal L}^i(x;\gamma) v = \Lambda v, \quad x\in\mathbb{R},
\]
has an eigenvalue $\Lambda =0$ if and only if one of the following two
conditions holds 
\begin{romannum}
\item $D_{xx}\phi^i_\pi(x;a;\gamma)$ is continuous at $x=0$, i.e.,
  $\phi^i_\pi(0;a;\gamma)=k\pi-a^2\frac{\gamma}{12}+{\cal O}(a^4)$,
  for some $k\in \mathbb{Z}$;
\item $D_x\phi_\pi^i(0;a;\gamma) =0$ and there are some $x_\pm$, with $\mathrm{sgn}(x_\pm)= \pm1$, such
  that $D_x\phi_\pi^i(x_\pm;a;\gamma)\neq 0$.
\end{romannum}
\end{lemma}
\begin{proof}
As the proof of Lemma~\ref{lem:zero} is based on the fact that the
derivative of the semifluxon is a solution of the linearized system
for $x\neq 0$, we can follow the same arguments to prove this
lemma. Again this leads to two conditions that either $\phi^i_{xx}$ is
continuous at $x=0$ or the second condition as stated above. 

In order to determine when $\phi^i_{xx}$ is continuous, we use the
static equation~(\ref{cont_appr_static}) and expand near $a=0$:
\[
\begin{array}{lll}
D_{xx}\phi^i_\pi(x;a;\gamma)&=&
\left(\sin(\phi^i_\pi(x;a;\gamma)+\theta(x))-\gamma\right)
\left(1-\frac{a^2}{12}\cos\widetilde\phi\right) \\
&&{} + 
\frac{a^2}{6}\sin\widetilde\phi
\left(\gamma\arcsin\gamma+\sqrt{1-\gamma^2}-\gamma\widetilde\phi-\cos\widetilde\phi\right)
+ {\cal O}(a^4).
\end{array}
\]
again with  $\widetilde\phi=\phi^i_\pi(x;0;\gamma)+\theta(x)$. The continuity of
$D_{xx}\phi^i_\pi$ at $x=0$ leads to the expression for
$\phi^i_\pi(x;a;\gamma)$ as given above. 
\end{proof}

At $\gamma=\gammacr(a)$, the stable manifold of the
$\pi+\arcsin\gamma$ and the homoclinic connection at $\arcsin\gamma$
are tangent, implying that $D_{xx}\phi^i_\pi(x;a;\gamma)$ is
continuous at $x=0$. Thus the first condition of the lemma is satisfied at
$\gamma=\gammacr(a)$ for $i=1,3$. For the same reasons as before, the
second condition is never satisfied. 

Since $\Lambda=0$ is an eigenvalue of the linearized
operator~${\cal L}^i(x;a;\gamma)$ if and only if $\gamma=\gammacr(a)$,
the sign of the eigenvalues of ${\cal L}^i(x;a;\gamma)$ will not
change. Thus the behavior of the eigenvalues near $\gamma=0$ will
again determine the stability of the semifluxons.

For $\gamma=0$ and $\theta=0$, the sine-Gordon equation with a
perturbation due to the lattice spacing has a heteroclinic orbit
connecting~$0$ and~$2\pi$. As before, the heteroclinic orbit will play
an important role in determining the stability of the semifluxons for
small values of~$\gamma$.  For small values of the lattice
spacing~$a$, we can approximate this heteroclinic orbit up to order
$a^2$ by using the $2\pi$-fluxon $\pfl$ and its linearization.
\begin{lemma}
  Let $\pfl^a(x)$ denote the heteroclinic orbit of the sine-Gordon
  equation with a perturbation due to the lattice spacing
  (i.e.,~(\ref{cont_appr}) with $\theta\equiv0$ and $\gamma=0$). For
  the lattice spacing~$a$ small, we have for the symmetric (i.e.,
  $\pfl^a(0)=\pi$) heteroclinic connection $\pfl^a(x)$
\begin{equation}
\pfl^a(x)=\pfl(x)+a^2\phi_a(x)+{\cal O}(a^4),
\end{equation}
where
\begin{equation}
\phi_a(x)=-\frac1{12}\frac{-3\sinh x+x\cosh x}{\cosh^2x}.
\label{phia}
\end{equation}
This approximation is valid, uniform in $x\in\mathbb{R}$.
\end{lemma}

\begin{proof}
  The spatially localized correction to the kink shape $\pfl(x)$ due
  to the perturbation term representing lattice spacing is sought in the
  form of perturbation series:
\[
\pfl^a(x)=\pfl(x)+a^2\phi_a(x)+\mathcal{O}(a^4).
\]
It is a direct consequence that $\phi_a(x)$ satisfies 
\begin{eqnarray}
\label{1stappr}
\mathcal{L}^1(x;0)\phi_a(x) = f(x)&=&
-{\textstyle\frac{1}{12}} \left[ 2\cos\pfl(x)\partial_{xx}\pfl(x)
\right.\\
&&\left.{}-\sin\pfl(x)(\partial_x\pfl(x))^2- 
\cos\pfl(x)\sin\pfl(x) \right],\nonumber
\end{eqnarray}
where $\mathcal{L}^1(x;0)$ is the linearized operator associated to
the fluxon, i.e., ${\cal L}^1(x;0)= D_{xx}-\cos\pfl(x)$.

Using the variation of constants method, we obtain the general
solution of (\ref{1stappr}), i.e.
\begin{equation}
\phi_a(x)=A(x)\,\sech\,{x}+B(x)\,({x}\,\sech\, {x}+\sinh {x}),
\label{1stappr_sol}
\end{equation}
where
\begin{eqnarray}
A(x) &=& A_0 + \frac{1}{24} \left[
2\ln\left(\frac{1-\cosh x-\sinh {x}}{\cosh {x}-1-\sinh {x}}\right) +
\frac{6\sinh {x}}{\cosh {x}}-\frac{4\sinh {x}}{\cosh^3{x}} +
\int_0^{{x}}{\frac{{\xi}f(\xi)}{\cosh {\xi}}d{\xi}}
\right],\nonumber\\ 
B(x) &= & B_0 - 
\frac{1}{24}\left[2+\frac{1}{\cosh^2{x}}-\frac{3}{\cosh^4{x}}\right].
\nonumber
\end{eqnarray}
The integration constant~$B_0$ is determined by the condition that
$\phi_a(x)$ is bounded, leading to $B_0=\frac{1}{12}$. The integration
constant~$A_0$ is determined by the requirement that $\pfl^a(0)=\pi$,
hence $\phi_a(0)=0$, giving that $A_0=0$.
\end{proof}

For $\gamma=0$, the static model~(\ref{cont_appr_static}) for a
$0$-$\pi$ Josephson junction with lattice spacing effects has both a
$\pi$- and a $3\pi$-kink solution. The $2\pi$-heteroclinic orbit found
above, can be used to derive approximations for those kinks.
\begin{lemma}\label{as_phi1}
  For $a$ small and $\gamma=0$, we have an explicit expression for the
  $\pi$- and $3\pi$-fluxon up to order ${\cal O}(a^2)$, respectively:
\begin{equation}
\begin{array}{lll}
  \phi^1_\pi(x;a;0) &=& \phi^1_\pi(x;0)+a^2\left\{
    \begin{array}{lll}
      -u^1_\pi(x-\ln(1+\sqrt2)),&\qmbox{for}&x<0\\
      u^1_\pi(-x-\ln(1+\sqrt2)),&\qmbox{for}&x>0
    \end{array}\right.\\
      \phi^2_{3\pi}(x;a;0) &=& \phi^2_{3\pi}(x;0)+a^2 \left\{
    \begin{array}{lll}
      -u^1_{3\pi}(x+\ln(1+\sqrt2)),&\qmbox{for}&x<0\\
      u^1_{3\pi}(-x+\ln(1+\sqrt2)),&\qmbox{for}&x>0
    \end{array}\right.\\
\end{array}
\label{13pi}
\end{equation}
where
$\phi^1_\pi(x;0)$ and $\phi^2_{3\pi}(x;0)$ are the $\pi$- resp.\ the
$3\pi$- fluxons as defined in~(\ref{sge2b}) and
\begin{eqnarray}
u^1_\pi(x) &=&\textstyle \frac1{12\cosh x} 
\left( \frac{3\sqrt2}2 -\frac12\ln(3-\sqrt2)+3\tanh x-x \right)
\nonumber\\
u^1_{3\pi}(x)&=&
\textstyle \frac1{12\cosh x} 
\left( -\frac{3\sqrt2}2 +\frac12\ln(3-\sqrt2)+3\tanh x-x \right)
.\nonumber
\end{eqnarray}
\end{lemma}

\subsection{Stability of type 1 semifluxon}

We will show that the type~1 wave $\phi^1_\pi(x;a;\gamma)$ is linearly
stable for small $a$ and $0\leq\gamma\leq\gammacr$ by analyzing the
largest eigenvalue of ${\cal L}^1(x;a;\gamma)$ for
$0\leq\gamma\leq\gammacr(a)$.

\begin{lemma}\label{lem:disc_1}
  For the lattice spacing parameter $a$ sufficiently small and
  $0\leq\gamma<\gammacr(a)$, the largest eigenvalue of ${\cal
    L}^1(x;a;\gamma)$ is strictly negative. For $\gamma=\gammacr(a)$,
  the operator ${\cal L}^1(x;a;\gammacr(a))$ has 0 as its largest
  eigenvalue. For $\gamma=0$, the largest eigenvalue decreases as $a$
  increases and is proportional to
  $-\frac14(\sqrt5+1)-0.0652a^2+\mathcal{O}(a^4)$.
\end{lemma}

\begin{proof}
  First we look at the stability of the $\pi$-kink at $\gamma=0$.
  Writing $v(x)=v^0(x)+a^2v^1(x)+{\cal O}(a^4)$ and
  $\Lambda=\Lambda_0+a^2\Lambda_1+{\cal O}(a^4)$ and expanding the
  eigenvalue problem for the stability of the $\pi$-kink
  $\phi_{\pi}^1(x;a;0)$
in a Taylor series, result in the following equations
\begin{equation}
\begin{array}{lll}
\left({\cal L}^1(x;0;0)-\Lambda_0\right)v^0(x)&=&0,\\
\left({\cal L}^1(x;0;0)-\Lambda_0\right)v^1(x) &=& 
\left(\Lambda_1-u^1_\pi(x)\,\sin(\phi_\pi^1(x;0)+\theta)\right)\,v^0(x)-g(x),
\end{array}
\label{EP_appr2}
\end{equation}
where $\mu =\sqrt{\Lambda_0+1}$,
$\Lambda_0=-\frac14(\sqrt5+1)$,
\[
\begin{array}{lll}
v^0(x) = \left\{
\begin{array}{ll}
e^{\mu(x-\ln(1+\sqrt2))} \, [\tanh(x-\ln(1+\sqrt2))-\mu] ,&\qmbox{for}x<0, \\
e^{\mu(-x-\ln(1+\sqrt2))} \, [\tanh(-x-\ln(1+\sqrt2))-\mu]
,&\qmbox{for}x>0 ,
\end{array}
\right.\\
g(x) = \frac{1}{12}\left[ 
2v^0_{xx}\Lambda_0 + v^0 + 2v^0_{xx}\cos\widetilde\phi(x) - 2\cos^2\widetilde\phi(x)v^0 
- 2\partial_{xx}(\phi_\pi^1(x;0))\sin\widetilde\phi(x)v^0
\right.\nonumber\\
\hspace{0.5cm}{} \left.{}
-2\partial_x\phi_\pi^1(x;0)\sin\widetilde\phi(x)v^0_x 
-(\partial_x\phi_\pi^1(x;0))^2\cos\widetilde\phi(x)v^0
-v^0\Lambda_0^2-2v^0\Lambda_0\cos\widetilde\phi(x)
\right],
\end{array}
\]
with again $\widetilde\phi(x)=\phi_\pi^1(x;0)+\theta(x)$ (see Lemma
\ref{lem:L1}).

The parameter value of $\Lambda_1$ is calculated by
solving~(\ref{EP_appr2}) for a bounded and decaying solution~$v^1(x)$.
The general solution can be derived by using the variation of constant
method because we have the homogeneous solutions of the equation. One
can also use the Fredholm theorem (see, e.g., \cite{scot99}), i.e.\
the sufficient and necessary condition for (\ref{EP_appr2}) to have
a solution $v^1\in H_2(\mathbb{R})$ is that the inhomogeneity is
perpendicular to the null space of the self-adjoint operator of ${\cal
  L}^1(x;0;0)$.
If $\langle\,,\,\rangle$ denotes an inner product in $H_2(\mathbb{R})$,
then this condition gives
\[
0 = \langle({\cal L}^1(x;0;0)-\Lambda_0)v^1,v^0\rangle = 
\langle\Lambda_1v^0-u^1_\pi
v^0\sin(\phi_\pi^1(x;0)+\theta)-g,v^0\rangle
\]
which implies that
\begin{equation}
\Lambda_1= \frac{3584(70\sqrt2(1+\sqrt5)-99(1+\sqrt5))}%
{24576(-70\sqrt{10}-350\sqrt2+495+99\sqrt5)}\approx-0.0652.
\label{lambda1pikink}
\end{equation}

Now assume that the operator ${\cal L}^1(x;\gamma)$ has a positive
eigenvalue~$\Lambda^1(\gamma)$ for some $0\leq\gamma<\gammacr(a)$. Since
$\Lambda$ depends continuously on $\gamma$, there has to be some
$0<\widehat\gamma<\gammacr(a)$ such that
$\Lambda^1(\widehat\gamma)=0$. However, from Lemma~\ref{lem:zero_a} it
follows that this is not possible. 
\end{proof}

\subsection{Instability of type 2 semifluxon}

In Lemma~\ref{lem:L2} we have seen that for $a=0$, the linearization
about the type~2 semifluxon has a strictly positive largest
eigenvalue. Also the limits of this eigenvalue for $\gamma\to0$ and
$\gamma\to\gamma^*$ are still strictly positive. Thus a small
perturbation associated with the lattice spacing can not stabilize the
type~2 semifluxons.

For completeness, we will consider the case $\gamma=0$.  In this
limit, the type~2 semifluxon can be seen as a concatenation of a
$3\pi$-kink and a $-2\pi$-kink. As before, the limit of the largest
eigenvalue for $\gamma\to0$ will be equal to the largest eigenvalue of
the $3\pi$-kink. We have seen that the largest eigenvalue of the
$3\pi$-kink at $\gamma=0$ and $a=0$ is strictly positive and the
following lemma shows that small lattice spacing effects increase this
eigenvalue. 
\begin{lemma}\label{lem:lattice_3pi}
  For the lattice spacing parameter $a$ sufficiently small, the largest
  eigenvalue of the linearization~${\cal L}^2(x;a;0)$ about the
  $3\pi$-kink~$\phi_{3\pi}^2(x;a;0)$ is strictly positive. Moreover, it 
  increases as $a$ increases and is proportional to
  $\frac14(\sqrt5-1)+0.0652a^2+\mathcal{O}(a^4)$.
\end{lemma}
\begin{proof}
  Note that the lowest order analytic expressions for the $\pi$- and
  the $3\pi$-kink differs only in the sign of the 'kink-shift'
  (see~(\ref{13pi})).  Because of
  this, we can follow the same steps as the proof of Lemma~\ref{lem:disc_1}. Writing
  the largest eigenvalue of a 3$\pi$-kink as
  $\Lambda=\Lambda_0+a^2\Lambda_1+\mathcal{O}(a^4),$ with
  $\Lambda_0=(\sqrt5-1)/4$ as has been calculated in
  Lemma~\ref{lem:L2}, we compute $\Lambda_1$ to be:
\begin{equation}
\Lambda_1 = 
\frac{3584(665857(\sqrt5-1)-470832\sqrt2(\sqrt5+1))}%
{24576(3329285-2354160\sqrt2-665857\sqrt5+470832\sqrt{10})}
\approx0.0652.
\label{l1_sfs2}
\end{equation}
\end{proof}

Thus up to order $\mathcal{O}(a^4)$ the lattice spacing effects
destabilize the 3$\pi$-kink.

Because a $2\pi$-fluxon in the 'ordinary' sine-Gordon equation can be
pinned by lattice spacing effects, one might expect to have a stable
3$\pi$-kink in the 0-$\pi$ sine-Gordon equation with larger lattice
spacing effects.  This is confirmed by numerical calculations in
section~\ref{3sec6}, see Figure~\ref{spectrum_3pi}.  If the
$3\pi$-kink is stable for $\gamma=0$, a stable type~2 semi-kink might
exist for $\gamma>0$ when the repelling force between the $3\pi$-kink
and the anti-fluxon is smaller than the energy to move a fluxon along
lattices. However, in section~\ref{3sec6} it will be shown numerically
that the type~2 semikink is unstable for all values of the lattice
spacing, see Figure~\ref{SF2}(b).

\subsection{Instability of type 3 semifluxon}

For $\gamma$ small or close to $\gammacr$, it has been shown in
Lemma~\ref{lem:L3} that the type~3 semifluxons are weakly unstable.
This opens the possibility that the perturbation term representing the
lattice spacing stabilizes the semifluxon. This is not the case
however.

\begin{lemma}\label{lem:L3:a}
  For small lattice spacing~$a$ and bias current
  $0<\gamma<\gammacr(a)$, the largest eigenvalue 
  of the linearization~${\cal L}^3(x;a;\gamma)$ about the type~3
  semifluxon~$\phi_\pi^3(x;a;\gamma)$ is strictly positive.  For
  $\gamma=\gammacr(a)$, the operator ${\cal L}^3(x;a;\gammacr)$ has~0
  as its largest eigenvalue. For $\gamma$ near zero and $a^2=\gamma
  \hat a^2$, the largest eigenvalue of ${\cal L}^3(x;a;\gamma)$ is
  $\La_\ast = \left(\frac\pi4+\frac7{180}\hat a^2\right) \, \gamma +
  \O(\gamma\sqgam)$.
\end{lemma}

\begin{proof}
  At $\gamma=\gammacr$, the
  solution~$\phi^3_\pi(x;a;\gammacr(a))=\phi^1_\pi(x;a;\gammacr(a))$.
  Hence from Lemma~\ref{lem:disc_1} it follows that the largest
  eigenvalue of the linearization about $\phi^3_\pi(x;a;\gammacr(a))$
  vanishes.  

  From Lemma~\ref{lem:L3} it follows that the largest eigenvalue of
  the linearization about $\phi^3_\pi(x;a;\gamma)$ is positive for
  $a=0$ and $0<\gamma<\gammacr$. Thus a small perturbation can not
  change the positive sign of the largest eigenvalue if $\gamma$ is
  not near $0$ or $\gammacr$. Now assume that a small perturbation
  would lead to a negative largest eigenvalue near $\gamma=0$ or
  $\gamma=\gammacr$. Then there has to be a zero eigenvalue near
  $\gamma=0$ or $\gamma=\gammacr$, but this is not possible according
  to Lemma~\ref{lem:disc_1}. Thus we can conclude that the largest
  eigenvalue is always positive. 

  To complete the proof, we will derive the asymptotic expression of
  the eigenvalue near $\gamma=0$. Since both $a$ and $\gamma$ are
  small, we relate those two parameters by writing $a^2=\gamma\hat
  a^2$. Now the approximation for the type~3 semifluxon can be written
  as
\[
\phi_\pi^3(x;\hat a\sqrt\gamma;\gamma) 
=\left\{
\begin{array}{lll}
\pfl(\widehat x) + \gamma\phi_1(\widehat x) + 
\gamma \hat a^2\phi_a(\widehat x) + \gamma^2 R_2(\widehat x;\gamma),
&&x<-L_\pi(\gamma)+x_1,\\
\pfl(-\widetilde x)+\gamma\phi_1(-\widetilde x) + 
\gamma \hat a^2\phi_a(-\widetilde x) + \gamma^2 R_2(-\widetilde x;\gamma),
&&-L_\pi(\gamma)+x_1<x<0,\\   
\pi+\pfl(-x-x_1) + {\cal O}(\gamma),&&x>0,
\end{array}\right.
\] 
where $\widehat x=x-x_1+2L_\pi(\gamma)$ and $\widetilde x = x-x_1$.
It can be shown that the shift $L_\pi(\gamma)$ does
not depend on $\hat a^2$ in lowest order, i.e., $L_\pi(\gamma) =
\frac12 |\ln \gamma| + \ln \frac{4}{\sqrt{\pi}} + \O(\sqrt\gamma)$. 

To find largest eigenvalue, we set again $\Lambda^3(\gamma) = \gamma
\La_1(0)$ and follow the steps in the proof of Lemma~\ref{lem:L3} with
some additional terms added to some expressions.

First, we consider the part of the approximation with
$x<-L_\pi(\gamma)+x_1$ or ${\widehat x}<L_\pi(\gamma)$. As before, we
drop the hat in $\widehat x$ in this part of the arguments. On
$(-\infty,L_\pi)$, the general solution of the eigenvalue problem of
the order ${\cal O}(\gamma)$ after expanding $\psi_{\rm approx}^1 =
\psi_0 + \gamma \psi_1$ is 
\begin{eqnarray}
\psi_1(x) &=&
\left[\frac\pi4 - \frac12 \La_1 \left(\ln\cosh x + 
\int_0^x \frac{\xi}{\cosh^2 \xi} d \xi\right)\right] 
\frac{1}{\cosh x}\nonumber\\ 
&&{} + \left[ \frac12 \La_1(0) + \frac12 \La_1 \tanh x\right]
\left(\frac{x}{\cosh x} + \sinh x\right)+
\frac12\left(\frac{d}{dx} \phi_1 + 
\hat a^2\frac{d}{dx} \phi_a \right)\nonumber\\
&&{} - \frac{e^x}{360(e^{2x}+1)^3}
\left[16\ln2+e^{2x}(32\ln2-295+60x)+30x+137 + 7e^{6x}\right.\nonumber\\
&&\hspace{1cm}\left.{}-16\ln(e^{2x}+1)(e^{2x}+1)^2+ e^{4x}(151+30x+16\ln2) \right]. \nonumber 
\end{eqnarray}
We note that the error term $|\psi(x) - \psi_{\rm appr}^1(x)| =
\gamma^2|S_2(x;\gamma)|$ is still at most $\O(\gamma)$ on $(-\infty,
L_\pi)$.

Next consider the second part of the approximation, i.e., $x$ between
$-L_\pi(\gamma)+x_1$ and 0 or $\widetilde x<-L_\pi(\gamma)$. Again, we
drop the tilde in $\widetilde x$. We scale $\psi(x)$ as $\gamma
\tpsi(x)$.  The linearization $\tpsi(x)$ along $\phi_{\rm appr}^2(x)$
on the interval $(-L_{\pi}, -x_1)$ must solve $\L \tpsi = O(\gamma)$.
Thus, at leading order 
\[
\tpsi(x) = \frac{\tilde{A}}{\cosh x} +
\tilde{B}\left(\frac{x}{\cosh x} + \sinh x\right).  
\]

The last part of the approximation of $\psi(x)$ on $x > 0$,
$\psi_{\rm appr}^3(x)$, is obtained by linearizing along $\phi_{\rm
  appr}^3(x)$ and by translating $x$ so that $x \in (x_1, \infty)$.
We also scale $\psi_{\rm appr}^3(x)= \gamma
\hat{\psi}(x)$. As $\hat{\psi}$ must be bounded for $x \to
\infty$, it follows that $\hat{\psi}(x) = \hat{A}/\cosh x +
\O(\gamma)$ for some $\hat{A} \in \mathbb{R}$.

Finally we have to connect all parts of the eigenfunction in a
$C^1$-fashion. This determines the values of 
$\Lambda_1(0)$, $\tilde{A}$, $\tilde{B}$, and $\hat{A}$ as
\[
\La_1(0) = \frac14 \pi+\frac7{180}\hat a^2,\quad \tilde{B} = -\La_1(0), 
\quad \tilde{A} = \frac14 \pi [\sqrt{2} + \log(\sqrt{2} - 1)]
,\qmbox{and} \hat{A} = \frac12 \pi \sqrt{2}, 
\]
thus $\Lambda_1(0)>0$, $\tilde{B}  < 0$, $\tilde{A} >0$ and $\hat{A} > 0$.
And we can conclude that the eigenvalue problem for the $\pi$-fluxon
$\pp3(x;\hat a\sqrt\gamma;\ga)$  has a positive largest eigenvalue 
\[
\La_\ast = \left(\frac\pi4+\frac7{180}\hat a^2\right)\gamma + \O(\gamma\sqgam).
\]
\end{proof}


\section{Semikinks in the weak-coupling limit}
\label{3sec5}

In this section we will consider the discrete $0$-$\pi$ sine-Gordon
equation~(\ref{dsg}) when the lattice parameter~$a$ is large. The time
independent version of~(\ref{dsg}) is well-known: when $\gamma=0$, it
corresponds to the so-called Standard or Taylor-Greene-Chirikov
map~\cite{chir79} and when $\gamma\neq0$, it is called the Josephson
map~\cite{nomu96}. Since we are interested in the case that the lattice
spacing~$a$ is large, we introduce the coupling parameter~$\eps$ as
$\eps=\frac1{a^2}$ and the equation becomes
\begin{equation}
  \label{dsg_eps}
  \ddot{\phi}_n - \eps\left[\phi_{n-1}-2\phi_n+\phi_{n+1}\right] =
-\sin(\phi_n+\theta_n)+\gamma.
\end{equation}
When there is no coupling, i.e. $\eps=0$, it can
be seen immediately that there are infinitely many steady state
solutions:
\[
\phi_n = \left\{ 
\begin{array}{llll}  
\cos(k_n\pi)\arcsin\gamma+k_n\pi,&&n=0,-1,-2,\dots\\
\cos(k_n\pi)\arcsin\gamma+(k_n+1)\pi,&&n=1,2,3,\dots,
\end{array}\right. 
\]
where $k_n$ is an integer. The only monotone semi-kink is the solution
with $k_n=0$ for $n\in\mathbb{Z}$, thus it is natural to identify
this semikink with the type~1 semikink. However, it is less clear
which solution would correspond to the type~2 and type~3 semi-kinks.
Possible candidates for the type~2 wave are solutions for which there
is some $N\in\mathbb{N}$ such that $k_n=0$ for $n\leq0$ and $n\geq N$
and $k_n=1$ for $0<n<N$. Similarly, candidates for the type~3 wave are
solutions for which there is some $N\in\mathbb{N}$ such that $k_n=0$
for $n\leq-N$ and $n\geq 0$ and $k_n=1$ for $-N<n<0$. But there are
many other candidates involving combinations of $k_n=0$ or $k_n=1$ as
well.  If one starts with such a wave in the uncoupled limit, i.e., with
$\eps\ll1$ or $a\to\infty$ and uses continuation to follow this
wave in the discrete system~(\ref{dsg_eps}) towards $a=0$ or $\eps\to
\infty$, then it turns out that most waves end in a saddle-node
bifurcation. More details about the continuation can be found in
section~\ref{3sec6}.

In this section we will focus on the analytical study of the type~1
semi-kink for the coupling parameter~$\eps$ small (thus the lattice
spacing~$a$ large). We will denote this wave by
$\Phi^1_\pi(n;\eps;\gamma)$ and for $\eps=0$, we have 
\[
\Phi_\pi^1(n;0;\gamma)=
\left\{
\begin{array}{llll}
\arcsin\gamma,& n=0,-1,-2,\dots\\
\pi+\arcsin\gamma,& n=1,2,3,\dots.
\end{array}
\right.
\]

The existence of the continuation of~(\ref{lsf1}) for small
coupling~$\eps$ is guaranteed by the following lemma.
\begin{lemma}\label{lem:ES}
  The steady state solution $\Phi^1_\pi(n;0;\gamma)$, representing the
  semifluxon of type 1 in the uncoupled limit $\eps=0$, can
  be continued for $\eps$ small and $\gamma<1$. It is given by 
\begin{equation}
\Phi^1_\pi(n;\eps;\gamma) = 
\left\{\begin{array}{llll}
  \arcsin\gamma + {\cal O}(\eps^2),&& n\leq -1;\\
\arcsin\gamma + \eps \, \frac\pi{\sqrt{1-\gamma^2}} + {\cal O}(\eps^2),
&& n=0;\\
\pi + \arcsin\gamma - \eps \,\frac\pi{\sqrt{1-\gamma^2}} + {\cal O}(\eps^2), 
&& n=1;\\
\pi + \arcsin\gamma + {\cal O}(\eps^2),&& n\geq 2.
\end{array}\right.
\label{lsf1}
\end{equation}

For $\gamma$ close to one, we write
$\gamma=1-\eps\widetilde{\gamma}$.
If $\widetilde\gamma > \pi$, then the type~1 solution is:
\begin{equation}
\Phi^1_\pi(n;\eps;1-\eps\widetilde\gamma)=
\left\{
\begin{array}{llll}
\frac\pi2-\sqrt{\eps}\sqrt{2\widetilde{\gamma}} +
{\cal O}(\eps),&& n\leq-1;\\
\frac\pi2-\sqrt{\eps}\sqrt{2(\widetilde{\gamma}-\pi)} +
{\cal  O}(\eps),&& n=0;\\ 
\frac{3\pi}2-\sqrt{\eps}\sqrt{2(\widetilde{\gamma}+\pi)} + 
{\cal O}(\eps),&& n=1;\\
\frac{3\pi}2-\sqrt{\eps}\sqrt{2\widetilde{\gamma}}+
{\cal O}(\eps),&& n\geq2;\\
\end{array}
\right.
\label{lsf1_g}
\end{equation}
From~(\ref{lsf1_g}) we obtain
the critical bias current for the existence of static semifluxon as
\begin{equation}
\gammacr=1-\eps\pi+{\cal O}(\eps^2).
\label{ga2}
\end{equation}
\end{lemma}
\begin{proof}
  The existence proof for $\gamma<1$ follows from the implicit
  function theorem as given in \cite[Theorem 2.1]{mack95} or
  \cite[Lemma 2.2]{peli05}.

  For the case $\gamma=1-\eps\widetilde\gamma$, the implicit function
  theorem as presented in the references above can not be applied
  immediately. However, after some manipulations, the implicit
  function theorem can be applied again. First we substitute
  into the steady state equation $\gamma=1-\eps\widetilde\gamma$ and
  $\Phi=\Phi_0 + \sqrt\eps\widetilde\Phi$, where
  $\Phi_0(n)=\frac\pi2$, for $n\leq 0$ and $\Phi_0(n)=\frac{3\pi}2$,
  for $n\geq1$. This gives the following equations:
\[
\begin{array}{llll}
0 &=& \frac{\cos(\sqrt\eps\widetilde\Phi(n)) -1}\eps 
+\widetilde\gamma -\sqrt\eps
[\widetilde\Phi(n-1)-2\widetilde\Phi(n)+\widetilde\Phi(n+1)] 
=:\widetilde F_n(\widetilde\Phi,\eps)   ,\quad n\neq 0,1\\
0 &=& \frac{\cos(\sqrt\eps\widetilde\Phi(0)) -1}\eps 
+\widetilde\gamma -\sqrt\eps
[\widetilde\Phi(-1)-2\widetilde\Phi(0)+\widetilde\Phi(1)] -\pi 
=:\widetilde F_0(\widetilde\Phi,\eps)   ,\quad n= 0\\
0 &=& \frac{\cos(\sqrt\eps\widetilde\Phi(1)) -1}{\eps} 
+\widetilde\gamma -\sqrt\eps
[\widetilde\Phi(0)-2\widetilde\Phi(1)+\widetilde\Phi(2)] +\pi 
=:\widetilde F_1(\widetilde\Phi,\eps)   ,\quad n= 1
\end{array}
\]
Using that $\lim_{\eps\to0}\frac{\cos(\sqrt\eps\widetilde\Phi(n))
  -1}\eps = -\frac12(\widetilde\Phi(n))^2$, the definitions for
$\widetilde F$ can be smoothly extended to $\eps =0$ too. The
equations for $\eps=0$ become
\[
\widetilde\Phi^2(n)=2\widetilde\gamma, \,\, n\neq0,1;\quad
\widetilde\Phi^2(0)=2(\widetilde\gamma-\pi); \qmbox{and}
\widetilde\Phi^2(1)=2(\widetilde\gamma+\pi).
\]
For $|n|$ large, wave should be asymptotic to the center point of the
temporal dynamics, hence $\widetilde\Phi(n)=-\sqrt{2\widetilde\gamma}$
for $|n|$ large. So for $\widetilde\gamma \geq \pi$, there are two
monotone semi-kinks (recall that the full semi-kink
is given by $\Phi_0+\sqrt\eps\widetilde\Phi$):
\[
\widetilde\Phi^\pm(n;0;\widetilde\gamma) = \left\{
  \begin{array}{llll}
    -\sqrt{2\widetilde\gamma}, && n\leq -1;\\
    \pm\sqrt{2(\widetilde\gamma-\pi)}, && n=0;\\
    -\sqrt{2(\widetilde\gamma+\pi)}, && n=1;\\
    -\sqrt{2\widetilde\gamma}, && n\geq 2.
  \end{array}\right.
\]
Note that the $\pm$-solutions collide for $\widetilde \gamma=\pi$. The
linearization $D\widetilde F(\widetilde\Phi^\pm,0)$ is invertible for
$\widetilde\gamma>\pi$, hence the implicit function theorem can 
be applied again and we have the existence of monotone semi-kinks
$\Phi_0(n)+\sqrt\eps\widetilde\Phi^\pm(n,\eps,\widetilde\gamma)$. In
analogue with the continuum case, the type~1 wave is the one that has
the discontinuity at the lowest value of the phase.
The critical bias current for the existence of a
static lattice semifluxon follows immediately from the arguments
above.
\end{proof}

The two $\pm$-solutions near $\gammacr$ as derived above in the
proof are like the type~1 and type~3
semi-fluxons near~$\gammacr$ in the PDEs studied in the previous two
sections. So in analogue to those PDEs, we can define for
$\widetilde\gamma>\pi$ 
\[
\Phi^1_\pi(n;\eps;1-\eps\widetilde\gamma) = 
\Phi_0(n)+\sqrt\eps\widetilde\Phi^-(n;\eps;\widetilde\gamma) 
\qmbox{and} \Phi^3_\pi(n;\eps;1-\eps\widetilde\gamma) = 
\Phi_0(n)+\sqrt\eps\widetilde\Phi^+(n;\eps;\widetilde\gamma) .
\]
where $\Phi_0$ and $\widetilde\Phi^\pm$ are as in the proof
above. Thus we get
\begin{equation}\label{eq:phi_discr_13}
\Phi^{1/3}_\pi(n;\eps;1-\eps\widetilde\gamma)=
\left\{
\begin{array}{llll}
\frac\pi2-\sqrt{\eps}\sqrt{2\widetilde{\gamma}} +
{\cal O}(\eps),&& n\geq-1;\\
\frac\pi2\mp\sqrt{\eps}\sqrt{2(\widetilde{\gamma}-\pi)} +
{\cal  O}(\eps),&& n=0;\\ 
\frac{3\pi}2-\sqrt{\eps}\sqrt{2(\widetilde{\gamma}+\pi)} + 
{\cal O}(\eps),&& n=1;\\
\frac{3\pi}2-\sqrt{\eps}\sqrt{2\widetilde{\gamma}}+
+{\cal O}(\eps),&& n\geq2;
\end{array}
\right.
\end{equation}

The spectral stability of $\Phi^i_\pi(n;\eps;\gamma)$ is obtained by
substituting $\phi_n=\Phi^i_\pi(n;\eps;\gamma)+v_ne^{\lambda t}$ in the
model equation~(\ref{dsg}). Disregarding the higher order terms in
$v_n$ gives the following eigenvalue problem
\begin{equation}
L^i(\eps;\gamma)\nu=\Lambda \nu,
\label{DEVP}
\end{equation} 
where $\Lambda=\lambda^2$, $\nu=(\dots,v_{-1},v_0,v_1,\dots)^T$ and
$L^i(\eps;\gamma)$ is the linear discrete operator
\[\begin{array}{lll}
&&L^i(\eps;\gamma)=
\left(\begin{array}{ccccccc}
\ddots & \ddots & \ddots    &          &           &         & 0\\
       & \eps      & -2\eps-A_{-1}  & \eps        &           &         &\\
       &        &    \eps      & -2\eps-A_{0}  & \eps         &         &\\
       &        &           &  \eps       &  -2\eps-A_{1}  &  \eps       &\\
    0  &        &           &           & \ddots   & \ddots  & \ddots
\end{array}\right)\\
&&A_n=\cos\left(\Phi^i_\pi(n;\eps;\gamma)+\theta_n\right),\quad n\in\mathbb{Z}.
\end{array}
\] 
This operator plays a similar role as the differential operator ${\cal
  L}^i(x;\gamma)=D_{xx} - \cos(\phi^i_\pi(x;\gamma)+\theta(x))$ in
section~\ref{3sec3}. The eigenvalue problem is an infinite dimensional
matrix problem for a real and symmetric matrix. Thus the eigenvalues
must be real.

In the discrete case, the continuous spectrum of semikinks is bounded.
The spectrum is obtained by substituting 
$v_n=e^{-ikn}$ in~(\ref{DEVP}) with
$J_n^i=-2\eps-{\sqrt{1-\gamma^2}}$ from which one obtains the
following dispersion relation for such linear waves
\begin{equation}
  \Lambda=-\left(\sqrt{1-\gamma^2}+4\eps\sin^2(\textstyle\frac{k}{2})\right).
\end{equation}
Thus the continuous spectrum consists of the intervals $\pm
i[\sqrt[4]{1-\gamma^2}\,,\,\sqrt{\sqrt{1-\gamma^2}+4\eps}]$ (recall
that $\Lambda=\lambda^2$).

In the following two lemmas we will show that all eigenvalues of the
linearization~$L^1(\eps;\gamma)$ are negative for $\eps$ small. Thus for
$\eps$ small, the type~1 wave is always stable. For
$\gamma=1-\eps\widetilde\gamma$ and $\widetilde\gamma>\pi$, it can be
shown that the linearization~$L^3(\eps;1-\eps\widetilde\gamma)$ has a
positive eigenvalue for $\eps$ small. Hence the type~3 wave is
unstable for $\gamma$ near $\gammacr$ and with $\eps$ small.  

\begin{lemma}
\label{lem:SF1}
For $0<\gamma<1$ and $\eps$ small, the largest eigenvalue of the 
operator $L^1(\eps;\gamma)$ is negative up to ${\cal
  O}(\eps^2)$. 
  
\end{lemma}
\begin{proof}
  The eigenvalue problem to calculate the stability of the
  monotone discrete $\pi$-kink $\phi^1_\pi(n;\eps;\gamma)$,
  ${n\in\mathbb{Z}}$ is given by~(\ref{DEVP}) with $i=1$.  Slightly
  modifying Baesens, Kim, and MacKay \cite{baes98}, the spatially
  decaying solution that corresponds to an eigenvalue of the
  above eigenvalue problem, can be approximated by
\begin{equation}
v_n=\left\{\begin{array}{llll}
c\ell^{-n},\,n\leq0,\\
\hat c\,c \ell^{n-1},\,n\geq1,
\end{array}\right.
\label{v_n}
\end{equation}
for some $c$, $\hat c$ and $|\ell|<1$. The diagonal elements
in $L^1(\eps;\gamma)$ are $A_n=\sqrt{1-\gamma^2} +{\cal O}(\eps^2)$,
if $n\neq 0,1$, $A_0=\sqrt{1-\gamma^2}
-\eps\frac{\gamma\pi}{\sqrt{1-\gamma^2}} + {\cal O}(\eps^2)$ and
$A_1=\sqrt{1-\gamma^2} +\eps\frac{\gamma\pi}{\sqrt{1-\gamma^2}} +
{\cal O}(\eps^2)$.  Thus $A_0\neq A_1$, hence the need for two
parameters $c$ and $\hat c$ (modifying~\cite{baes98}, where
$\hat c=\pm1$ following from the symmetry $A_0=A_1$).

For small nonzero $\eps$, if we can match exponentially decaying
solutions (\ref{v_n}) on both sides from either end of the lattice to
a central site, then we obtain a candidate for an
eigenfunction. With~(\ref{DEVP}), the 
parameters $\ell$ and $\hat c$ will be determined up to
order~$\eps$. For $n\neq0,1$, the relation~(\ref{DEVP})
gives up to order~$\eps$
\begin{equation}
\label{n_inf}
\textstyle\Lambda=-\sqrt{1-\gamma^2}+\eps\left(\ell-2+\frac1\ell\right).
\end{equation}
At the central sites $n=0,1$ we get up to order~$\eps$
\begin{eqnarray}
\Lambda &=& \textstyle 
-\sqrt{1-\gamma^2}+\eps \frac{\gamma\pi}{\sqrt{1-\gamma^2}}
+\eps(\ell-2+\hat c);\label{n_0}\\
\Lambda &=& \textstyle
-\sqrt{1-\gamma^2}-\eps \frac{\gamma\pi}{\sqrt{1-\gamma^2}}
+\eps\left(\ell-2+\frac1{\hat c}\right).
\label{n_1}
\end{eqnarray}
Combining~(\ref{n_inf}),~(\ref{n_0}) and~(\ref{n_1}) shows that there
are two possible value for $\hat c$, being $\hat
c_{\pm}=-\frac{\pi\gamma}{\sqrt{1-\gamma^2}}
\pm\frac{\sqrt{1+(\pi^2-1)\gamma^2}}{\sqrt{1-\gamma^2}}+{\cal O}(\eps)$ and
leads to the eigenvalue $\Lambda$ and the decay exponent $\ell$ as a
function of $\eps$ and ${\gamma}$, i.e.
\begin{eqnarray}
\ell_\pm&=&
\pm\frac{\sqrt{1-\gamma^2}}{\sqrt{1+(\pi^2-1)\gamma^2}}
+{\cal O}(\eps),\label{ell}\\
\Lambda_\pm&=&
-\sqrt{1-\gamma^2} +\frac\eps{\ell_\pm}\left(\ell_\pm-1\right)^2  +
{\cal O} (\eps^2). 
\label{Lmb}
\end{eqnarray}

General Sturm-Liouville theory states that a critical eigenfunction
that corresponds to the largest eigenvalue of a continuous eigenvalue
problem does not vanish, except probably at $x\to\pm\infty$.  This
theorem can also be extended to a discrete eigenvalue problem such
that the most critical eigenvector does not have sign
changes~\cite{atki64}. Thus, if we have a solution of the form
(\ref{v_n}) with $\ell>0$, then it is the critical eigenvector.

From~(\ref{ell}), we see that $\ell_+>0$,
  thus the largest eigenvalue $\Lambda_+$ from~(\ref{Lmb}) is in
  the gap between zero and the interval associated with the continuous
  spectrum, i.e.\ $\Lambda_+<0$.
\end{proof}

\begin{remark} \rm
  \label{rem:ES}  
  From the details in the proof, note that weak coupling with strong
  bias current leads to one additional eigenvalue associated
  with~$\ell_-$, where $\ell_-<0$ and $|\ell_-|<1$. This indicates
  that the eigenvector of the form (\ref{v_n}) is localized but has
  out of phase configuration, i.e.\ has infinitely many sign changes.
  This is a typical characteristic of a 'high-frequency' eigenvalue
  which is confirmed by the fact that $\Lambda_-$ is indeed smaller
  than the phonon band. The presence of a high-frequency eigenvalue of
  a kink was previously reported by Braun, Kivshar, and Peyrard
  \cite{brau97} in their study on the Frenkel-Kontorova model with the
  Peyrard-Remoissenet potential \cite{peyr82}.
\end{remark}

For $\gamma$ close to~1, i.e $\gamma=1-\eps\widetilde\gamma$, with
$\widetilde \gamma>\pi$, both the type~1 and the type~3 wave as given
in~(\ref{eq:phi_discr_13}) can be analyzed. In the following lemma 
we also show that both types have a high-frequency eigenvalue.
\begin{lemma}
\label{lem:SF1_gcr}
For $\gamma=1-\eps\tilde{\gamma}$, with $\widetilde\gamma>\pi$,
the largest eigenvalue of the operator~$L^1(\eps;1-\eps\widetilde\gamma)$ 
is strictly negative and the largest eigenvalue of the operator~$L^3(\eps;1-\eps\widetilde\gamma)$ 
is strictly positive. 
\end{lemma}

\begin{proof}
As before, we write for an eigenfunction
\[
v_n=\left\{\begin{array}{llll}
c\ell^{-n},\,n\leq0,\\
\hat c\,c \ell^{n-1},\,n\geq1,
\end{array}\right.
\]
for some $c$, $\hat c$ and $|\ell|<1$ and we substitute this in the
eigenvalue problem, which leads to the equations
\begin{eqnarray}
\Lambda &=& -\sin\left(\sqrt{\eps2\widetilde{\gamma}}+{\cal
     O}(\eps)\right) + \eps(1/\ell-2+\ell);\label{p1}\\
\Lambda &=& \mp\sin\left(\sqrt{2\eps(\widetilde{\gamma}-\pi)}+{\cal
     O}(\eps)\right) + \eps(\ell-2+\hat c);\label{p2}\\
\Lambda &=& -\sin\left(\sqrt{2\eps(\widetilde{\gamma}+\pi)}+{\cal
     O}(\eps)\right) + \eps(\ell-2+1/{\hat c}).\label{p3}
\end{eqnarray}
where $\mp$-sign in the second equation is a minus sign for the
eigenvalue problem associated with the type~1 wave and the other sign
in case of the type~3 wave.
Again, by subtracting~(\ref{p3}) from~(\ref{p2}), we get a quadratic
equation for $\hat c$, with two solutions, one of order
$\frac{1}{\sqrt\eps}$ and one of order $\sqrt\eps$ (which is easiest found by
writing the equation as a quadratic equation for $\frac1{\hat c}$):
\[
\begin{array}{llll}
\frac1{\hat c_1} &=& \frac 1{\sqrt\eps }
\left(\sqrt{2(\widetilde{\gamma}+\pi)} \mp
\sqrt{2(\widetilde{\gamma}-\pi)} + {\cal O}(\sqrt\eps)\right);\\
\hat c_2 &=& \frac1{\sqrt\eps}\left(-\sqrt{2(\widetilde{\gamma}+\pi)} \pm
\sqrt{2(\widetilde{\gamma}-\pi)} + {\cal O}(\sqrt\eps\right).
\end{array}
\]
Combining~(\ref{p1}) and~(\ref{p2}) resp.~(\ref{p3}) and using the two
expressions above gives that in both cases $\ell$ is of order~$\sqrt\eps$
and given by 
\[
\begin{array}{llll}
\frac1{\ell_1} &=& \frac 1{\sqrt\eps}
\left(\sqrt{2\widetilde{\gamma}} \mp
\sqrt{2(\widetilde{\gamma}-\pi)} + {\cal O}(\sqrt\eps)\right);\\
\frac1{\ell_2} &=& \frac1{\sqrt\eps}\left(
\sqrt{2\widetilde\gamma}-\sqrt{2(\widetilde{\gamma}+\pi)} 
+ {\cal O}(\sqrt\eps\right).
\end{array}
\]
Finally, substitution into~(\ref{p1}) shows that
\begin{equation}
\begin{array}{llll}
\Lambda_1 &=& \mp \sqrt\eps\sqrt{2(\widetilde{\gamma}-\pi)} +{\cal 
  O}(\eps);\\
\Lambda_2 &=& -\sqrt\eps\sqrt{2(\widetilde{\gamma}+\pi)} +{\cal
  O}(\eps).
\end{array}
\label{ell3}
\end{equation}
The eigenvalue that corresponds to $\ell>0$ is $\Lambda_1$. 

So clearly the largest eigenvalue $\Lambda_1$ is negative in case of the type~1 wave
and is positive in case of the type~3 wave. 

In addition, the operator~$L^1(\eps;1-\eps\widetilde\gamma)$ and 
$L^3(\eps;1-\eps\widetilde\gamma)$ have the same high-frequency eigenvalue 
$\Lambda_2$ (up to order~$\eps^2$).
\end{proof}

The proofs of lemmas~\ref{lem:SF1} and~\ref{lem:SF1_gcr} show the
presence of a high-frequency eigenvalue for a semi-kink in case the
bias current is not small. In the following we will show that
the eigenvalue appears when the bias current is larger than
$\sqrt\eps+{\cal O}(\eps)$.

Because we do not have an analytic expression for the type 2 and type 3
semi-kink in the small forcing limit, the analysis is done
only for the type 1 semi-kink.


\begin{lemma}
\label{lem:SF1_hf}
There is a critical value $\gamma_{\rm hf}$, $\gamma_{\rm hf} = 
\sqrt{\eps} + {\cal O}(\eps)$ such that for all 
$\gamma \in (\gamma_{\rm hf}, \gamma_{\rm cr})$ the 
operator~$L^1(\eps;\hat\gamma)$ has a high frequency eigenvalue 
that up to ${\cal O}(\eps^3)$ is attached to the lowest boundary 
of the continuous spectrum. The corresponding eigenvector is 
localized and changes sign between any two adjacent sites. 
\end{lemma}

The appearance of this eigenvalue and the structure of its 
eigenvector is checked numerically in section 6. 

\begin{proof}
Again, we write for an eigenvector
\[
v_n=\left\{\begin{array}{llll}
c\ell^{-n},\,n\leq0,\\
\hat c\,c \ell^{n-1},\,n\geq1,
\end{array}\right.
\]
for some $c$, $\hat c$ and $|\ell|<1$ and we substitute this in the
eigenvalue problem. We first consider $\gamma = \sqrt{\eps}\hat{\gamma}$.
This gives $A_n=1-\frac{\eps\hat\gamma^2}2-\frac{\eps^2\hat\gamma^4}8 +{\cal O}(\eps^{5/2})$, if $n\neq 0,1$, $A_0=1-\frac{\eps\hat\gamma^2}2-\eps^{3/2}\pi\hat\gamma-\eps^2(\frac{\hat\gamma^4}8+\frac{\pi^2}2)+{\cal O}(\eps^{5/2})$ and $A_1=1-\frac{\eps\hat\gamma^2}2+\eps^{3/2}\pi\hat\gamma-\eps^2(\frac{\hat\gamma^4}8+\frac{\pi^2}2)+{\cal O}(\eps^{5/2})$. 

Using the same procedures, this implies $\hat c_\pm = -\sqrt\eps\pi\hat\gamma-\eps^{3/2}\hat\gamma^3\pi
\pm\sqrt{\pi^2\hat\gamma^2\eps+2\pi^2\hat\gamma^4\eps^2+1}$ and 
\[
\frac1{\ell_\pm} = \eps\pi^2\pm2\sqrt{\pi^2\hat\gamma^2\eps+2\pi^2\hat\gamma^4\eps^2+1}
 = \pm 1 +\frac{\eps\pi^2(1\pm\hat\gamma^2)}2 +{\cal O}(\eps^2).
\]



For $\hat{\gamma} > 1$ there are two solutions $|\ell_{\pm}|<1$;
$\ell_+ > 0$ corresponds to the largest eigenvalue and also
exists for $\hat{\gamma} \leq 1$ (Lemma \ref{lem:SF1});
$\ell_-< 0$, so that its associated eigenvector $v_n$ indeed 
changes sign between any two adjacent sites.

The value $\gamma_{\rm hf} = \sqrt{\eps} + {\cal O}(\eps)$ indicates
the appearance of this high frequency eigenvalue from the continuous
spectrum. It follows from a straightforward analysis that this
eigenvalue exists, i.e. $\ell_- \in (-1,0)$ exists, for all $\gamma
\in (\gamma_{\rm hf}, \gamma_{\rm cr})$, and the corresponding
eigenvalue is
$\Lambda_-=-1+(\frac{\hat\gamma^2}{2}-4)\eps+\hat\gamma^4/8\eps^2+{\cal
  O}(\eps^3)$.  Up to ${\cal O}(\eps^3)$ this eigenvalue is nothing
else but the lower boundary of the continuous spectrum.
\end{proof}


\section{Numerical computations of the discrete system}
\label{3sec6}
To accompany our analytical results, we have used numerical
calculations. For that purpose, we have made a continuation program
based on Newton iteration technique to obtain the stationary kink
equilibria of~(\ref{dsg}) and~(\ref{bc_dsg}) and an eigenvalue problem
solver in \textsc{MATLAB}. To start the iteration, one can choose
either the continuum solutions discussed in section~\ref{3sec3}, i.e.,
the case where the lattice spacing parameter $a=0$, or trace the
equilibria from the uncoupled limit $\eps=0$ ($a\to\infty$) as
discussed in the previous section.  We use the number of computational
sites $2N=800$ for parameter values of $a=0.05$ or larger ($\eps=20$
or lower).

\subsection{Stability of type 1 lattice semifluxon}

\begin{figure}[tb!]
\begin{center}
\vspace{2cm}
\end{center}
\caption{(a) Two lattice semifluxons of type 1 with no bias current
  ($\gamma=0$) are plotted as a function of the lattice index, namely
  the kink for strong coupling with $\eps=100$ (equivalently, a very
  small lattice spacing $a=0.1$) ($-\ast-$), i.e.\ close
  to~(\ref{sge2a}) and the kink for weak coupling with $\eps=\frac14$
  (equivalently, a large lattice spacing $a=2$) ($-$o$-$). 
  (b) Numerically computed spectrum of a lattice
  semifluxon against the lattice spacing parameter~$a$ with $\gamma=0$. We
  used the number of sites $2N=300$. We zoom in the plot of spectra
  around -1 for clarity. The bold-solid-line is the calculated
  approximate function for the point spectrum using perturbation
  theory for $a$ small resp.\ for $\eps$ small.}
\label{latsf}
\end{figure}

The type 1 lattice semifluxon
$\Phi^1_\pi(n;\eps;0)$, $n\in\mathbb{Z}$ has been studied 
analytically both in the strong coupling limit ($a \ll 1$, 
or $\eps \gg 1$) and the weak coupling limit ($a \gg 1$, 
$\eps \ll 1$). In Figure~\ref{latsf}(a), $\Phi^1_\pi(n;\eps;0)$ 
is plotted for two different values of the coupling 
parameter~$\eps$. For a given value of~$\eps$, one can use as 
initial guess in the numerical procedure either a solution from the continuous limit~(\ref{sge2a}) or from
the uncoupled limit that has been discussed in the preceding
sections.

In Figure~\ref{latsf}(b), we present the numerically calculated
spectrum of the type~1 semifluxon with $\gamma=0$ as a function of the
lattice spacing parameter.  The approximate largest
eigenvalue~(\ref{lambda1pikink}), derived for~$a$ small, and the one
derived in Lemma~$\ref{lem:SF1}$ for $a$ large, are in a good
agreement with the numerically obtained largest eigenvalue. Any
eigenvalue below $\Lambda=-1$ belongs to the continuous spectrum. For
$a$ close to zero we do not see dense spectra because of the number of
sites we used. By increasing the sites-number we will obtain a more dense
spectrum.

There is only one eigenvalue outside the phonon bands, the largest 
eigenvalue as studied in
Lemma~\ref{lem:SF1}. This is in contrast to the case of an ordinary
lattice $2\pi$-kink~\cite{kevr00,kivs98} where there is an internal
mode bifurcating from the phonon band when the parameter~$a$
increases. 

If Fig.\ \ref{latsf}(b) shows the spectrum of the type 1 semifluxon 
as a function of the coupling parameter $\eps$ ($\eps=1/a^2$) for a fixed bias current $\gamma$, in Fig.\ \ref{SFC}
we present the numerically calculated spectrum of the type 1 lattice semifluxon 
as a function of $\gamma$ for a fixed $\eps$, $\eps=0.25$. 


Lemmas \ref{lem:SF1} and \ref{lem:SF1_hf} established the
existence of two eigenvalues (for $\eps$ small enough) for the
stability problem associated to the type 1 semifluxon, the largest
eigenvalue and an additional eigenvalue which bifurcates from the 
lower edge of the phonon band for bias current 
$\gamma > \gamma_{\rm hf}$. It follows from the numerical 
simulations that these are indeed the only two eigenvalues 
(Figure \ref{SFC}). For $\eps=0.25$, this minimum bias current 
$\gamma_{\rm hf}$ is approximately $0.466$. Interestingly, according 
to Lemma \ref{lem:SF1_hf} the bifurcation appears at
$\gamma_{\rm hf} = \sqrt\eps = 0.5$ at leading order in $\eps$.
This is in remarkably good agreement with the numerical result,
especially since the error is ${\cal O}(\eps)$ and $\eps = 0.25$. 

\begin{figure}[tb!]
  \begin{center}
\vspace{2cm}
\end{center}
  \caption{(a) Spectrum of the type~1 semifluxon as a function of the
  applied bias current~$\gamma$ for a value of the coupling constant
  $\eps=0.25$. The dashed line is theoretical prediction
  from~(\ref{Lmb}). 
  \newline
  In~(b) we zoom in on the spectrum around~$-1$ for clarity. The
  spectrum is normalized to the lower edge of the phonon band, i.e.\
  $\sqrt{1-\gamma^2}+4\eps$, such that the appearance of a high
  frequency eigenvalue can be seen clearly.} 
\label{SFC}
\end{figure}

  To picture the appearance of the high-frequency eigenvalue, all
  eigenvalues for the truncated $2N\times 2N$-matrix associated with
  $L^1(\eps;\gamma)$ are determined and the eigenvectors of the two
  largest eigenvalues and the two smallest eigenvalues are presented in
  Figures~\ref{ef_g0}-\ref{ef_g07} for various values of~$\gamma$ and
  a fixed~$\eps$. It can be observed that there is always a
  localized eigenfunction associated with the largest eigenvalue. In
  Figure~\ref{ef_g0} ($\gamma=0$), none of the other eigenvectors can
  be associated with localized eigenfunctions and in
  Figures~\ref{ef_g05} and~\ref{ef_g07}, the birth of the localized 
  eigenfunction associated with the smallest eigenvalue can be
  observed.

\begin{figure}[tb!]
  \begin{center}
\vspace{2cm}
\end{center}
\caption{
  The eigenvectors associated the two largest eigenvalues and the two
  smallest eigenvalues of the truncated $2N\times 2N$-matrix
  associated with~$L^1(\eps;\gamma)$ (the type 1 discrete semikink)
  for $\eps=0.25$ and $\gamma=0$. The results are shown with $2N=100$
  for clarity. Shown are the eigenvector of (a) the largest
  eigenvalue, (b) the second one, (c)-(d) the last two eigenvalues.
  There is only one eigenvalue for $L^1(\eps;\gamma)$ as there is only
  one localized eigenvector.
}
\label{ef_g0}
\end{figure}

\begin{figure}[tb!]
  \begin{center}
\vspace{2cm}\end{center}
\caption{The same as Fig.\ \ref{ef_g0} for $\gamma=0.5$.}
\label{ef_g05}
\end{figure}

\begin{figure}[tb!]
  \begin{center}
\vspace{2cm}\end{center}
\caption{
  The same as Figure~\ref{ef_g0} for $\gamma=0.7$. Note that there are
  now two localized eigenvectors shown in~(a) and~(d). The smallest
  eigenvalue associated with~(d) is $-1.7357$ while the lower edge of
  the phonon band is $-1.7141$. Note also that neighboring sites of
  the eigenvector in~(d) move out of phase indicating a high-frequency
  mode, contrary to the semikink's low-frequency mode in~(a).}
\label{ef_g07}
\end{figure}

If we keep increasing $\gamma$ further, then there is a critical applied bias current at which the largest eigenvalue becomes 0. Numerical computations show that this critical value is $\gammacr(\eps)$ above which static lattice semifluxons disappear. 

The critical bias current for the existence of a static type 1 lattice
semifluxon in the continuum limit and for a very weak coupling in the
discrete system has been discussed and analytical expressions were given in
sections~\ref{3sec3},~\ref{3sec4} and~\ref{3sec5},
respectively. In Figure~\ref{gamma_c}, the numerically calculated
critical bias current $\gammacr$ of the discrete system~(\ref{dsg}) as
a function of the lattice spacing~$a$ is presented. The approximate
functions, given in~(\ref{ga}) for small~$a$, and in~(\ref{ga2}) for
large~$a$ (small~$\eps$) are presented as dashed lines.

\begin{figure}[tb!]
\begin{center}
\vspace{2cm}
\end{center}
\caption{The critical bias current of a static $\pi$-kink as a
  function of the lattice spacing parameter~$a$. For~$\gamma$ above
  the critical current there is no static $\pi$-kink solution. The
  solid line is numerically obtained curve. Dashed lines are the
  theoretical predictions~(\ref{ga}) for $a\ll 1$ resp.~(\ref{ga2})
  for $a\gg 1$ ($\eps\ll1$).}
\label{gamma_c}
\end{figure}

\subsection{Instability of type 2 lattice semifluxon}

\begin{figure}[tb!]
  \begin{center}
\vspace{2cm}
\end{center}
\caption{Plot of the eigenvalues of a $3\pi$-kink as a function of the
  lattice spacing parameter~$a$. We zoom in the region with $a\ll1$
  where it shows that turning the lattice spacing on destabilizes the
  kink.  The dashed line depicts the analytically computed
  approximation~(\ref{l1_sfs2}) to the largest eigenvalue of the
  $3\pi$-kink.}
\label{spectrum_3pi}
\end{figure}

In the continuum models we have seen that for $\gamma$ small, the
instability of the type~2 semikink is mainly determined by the
instability of the $3\pi$-kink in the continuum models for $\gamma=0$ .
So we start this section by looking at the stability of the
$3\pi$-kink in the discrete model. We will denote the $3\pi$-kink by
$\Phi_{3\pi}^2(n;\eps;0)$, where as before  the coupling parameter
$\eps$ and the lattice spacing~$a$ are related by $\eps=\frac1{a^2}$.

Using our continuation program, we have followed a $3\pi$-kink
solution from the continuous limit $0<a\ll1$ up to the uncoupled
situation $\eps=0$ (i.e., $a=\infty$). 
We obtain that $\Phi_{3\pi}^2(n;0;0)$ is given by 
\begin{equation}
\Phi_{3\pi}^2(n;0;0)=
\left\{
\begin{array}{llll}
0,\quad n=-1,-2,\dots\\
2\pi,\quad n=0,\\
\pi,\quad n=1,\\
3\pi,\quad n=2,3,\dots.
\end{array}
\right.
\label{3pi_c0}
\end{equation}
Note that this discrete configuration is not monotonically increasing
as opposed to the continuum configuration, which is monotonic.  

In Figure~\ref{spectrum_3pi}, we present the numerically obtained
eigenvalues of a $3\pi$-kink as a function of the lattice spacing~$a$.
For small $a$, the largest eigenvalue is indeed increasing as is
predicted by the perturbation theory~(\ref{l1_sfs2}). As soon as the
lattice spacing is of order one, the largest eigenvalue decreases and
becomes zero at approximately $a=1.7521$.

After establishing that increasing the lattice spacing can stabilize
the $3\pi$-kink at $\gamma=0$, we continue by looking at the stability
of the $2\pi$-kink for $\gamma >0$. Interestingly, increasing the
lattice spacing does not stabilize a type~2 semikink for $\gamma>0$.
In Figure~\ref{SF2}, we show a plot of the type~2 semikinks for two
values of $\eps$ as well as a plot of the largest eigenvalue as a
function of $\eps$ for two particular values of $\gamma$, namely
$\gamma=0.01$ and $\gamma=0.1$.  We present the largest eigenvalue as
a function of the coupling~$\eps$ instead of the lattice spacing~$a$
as the eigenvalue changes most for small coupling (large lattice spacing). 
\begin{figure}[tb!]
\begin{center}
\vspace{2cm}
\end{center}
\caption{(a) Plot of a type 2 semikink with $\gamma=0.01$ for
  $\eps=100$ ($-\ast-$) and $\eps=40$ ($-$o$-$).  
\newline 
(b) Plot of the largest eigenvalue of a type~2 semikink as a
function of the coupling parameter~$\eps$. When $\eps=0$, the
eigenvalue converges to $\Lambda=\sqrt{1-\gamma^2}$.}
\label{SF2}
\end{figure}
From Figure~\ref{SF2} it follows that the solutions are unstable even
in the weak-coupling limit. This is interesting as in the limit for
$\gamma\to0$, the type~2 semikink can be seen as a concatenation of a
$3\pi$-kink and a $-2\pi$-kink. Both the $3\pi$-kink and $-2\pi$-kink
are stable for the coupling $\eps$ sufficiently small, while the
type~2 semikink turns out to be unstable. 

This instability issue can be explained by looking at the expression
of a type~2 semikink when it is uncoupled ($\eps=0$). For the two
particular choices of $\gamma$ above, we get from the simulations that the configurations
of the these semi-kinks are given by 
\begin{equation}
\Phi_\pi^2(n;0;0.01)=
\left\{
\begin{array}{llll}
0+\arcsin(0.01),& n\leq-1,\\
\pi-\arcsin(0.01),& n=0,\\
\pi+\arcsin(0.01),& n=1,\\
3\pi+\arcsin(0.01),& 2\leq n\geq8,\\
2\pi-\arcsin(0.01),& n=9,\\
\pi+\arcsin(0.01),& n\geq10,
\end{array}
\right.
\end{equation}
and
\begin{equation}
\Phi_\pi^2(n;0;0.1)=
\left\{
\begin{array}{llll}
0+\arcsin(0.1),& n\leq-1,\\
\pi-\arcsin(0.1),& n=0,\\
\pi+\arcsin(0.1),& n=1,\\
3\pi+\arcsin(0.1),& 2\leq n\geq6,\\
2\pi-\arcsin(0.1),& n=7,\\
\pi+\arcsin(0.1),& n\geq8.
\end{array}
\right.
\end{equation}

We see that there are two sites, namely $n=0$ and $n=9$ for
$\gamma=0.01$ and $n=0$ and $n=7$ for $\gamma=0.1$, where $\Phi$ takes
the value of an unstable fixed point of the discrete system~(\ref{dsg}).
Looking only at sites numbered $n=2$ to $n\to\infty$,
$\Phi_\pi^2(n;0;\gamma)$ can be viewed as a $-2\pi$ lattice kink
sitting on a site which is known to be unstable. If we look only at
sites numbered $n=6$ to $n\to-\infty$, $\Phi_\pi^2(n;0;\gamma)$
can be seen as a deformed $3\pi$ lattice kink (at site $n=0$, the
phase $\Phi$ takes the value $\pi$ instead of the value $2\pi$ as in
the $3\pi$-kink). Hence, it seems that coupling between the
two kinks due to the presence of a nonzero $\gamma$ is responsible for
the instability.

It has been discussed in the previous sections that there is a
critical bias current $\gamma^*$ for the existence of a type 2 lattice
semikink in the continuum models.  However, we did not numerically
calculate the critical bias current $\gamma^*(a)$ for discrete
system~(\ref{dsg}).

\subsection{Instability of type 3 lattice semifluxon}

\begin{figure}[tb!]
\begin{center}
\vspace{2cm}
\end{center}
\caption{(a) Plot of a type~3 semikink with $\gamma=0.01$ for
  $\eps=100$ ($-\ast-$) and $\eps=40$ ($-$o$-$).  
\newline 
(b) Plot of the largest eigenvalue of a type~3 semikink as a
function of the coupling parameter~$\eps$. When $\eps=0$, the
eigenvalue converges to $\Lambda=\sqrt{1-\gamma^2}$.}
\label{SF3}
\end{figure}

\begin{figure}[tbh!]
\begin{center}
\vspace{2cm}
\end{center}
\caption{
  Spectrum of the type 3 semifluxon as a function of the applied bias
  current~$\gamma$ for a value of the coupling constant~$\eps=0.25$.
  In~(b) and~(c) we zoom in near the phonon band for clarity. In~(b),
  the spectrum is normalized to the upper edge of the phonon band,
  i.e.\ $\sqrt{1-\gamma^2}$ and in~(c) it is normalized to the lower
  edge of the phonon band, i.e.\ $\sqrt{1-\gamma^2}+4\eps$. The
  disappearance of a high-frequency mode in the lower edge of the
  phonon band can  be clearly observed in~(c). The insets in~(b)
  and~(c) show the eigenfunctions of the two eigenvalues just above
  resp.\ just below the phonon band for $\gamma=0.73$. 
}
\label{specS3}
\end{figure}

In this section, we will consider the type~3 semikinks, which will be
denoted by $\Phi_\pi^3(n;\eps;\gamma)$. In Lemmas~\ref{lem:L3}
and~\ref{lem:L3:a} it has been shown that these kinks are unstable in the
continuum models for small or zero lattice spacing.

The largest eigenvalue of a lattice type 3 semifluxon for three
particular values of $\gamma$, i.e.\ $\gamma=0.01,\,0.1,\,0.55$, is
presented in Figure~\ref{SF3}. Even though in the limit for
$\gamma\to0$, a semifluxon of this type is a concatenation of a
$2\pi$-kink and a $-\pi$-kink which both can be stable in the discrete
system, the type~3 semikink is unstable for all parameter values from
the zero lattice spacing limit all the way to the zero coupling one.
The explanation is similar to the one for a type~2 semikink
discussed above.

Indeed, for the three particular choices of $\gamma$ above,
$\Phi_\pi^3(n;0;\gamma)$ is given by
\begin{equation}
\Phi_\pi^3(n;0;0.01)=
\left\{
\begin{array}{llll}
0+\arcsin(0.01),& n=-1,-2,\dots\\
\pi-\arcsin(0.01),& n=-6,\\
2\pi+\arcsin(0.01),& n=-5,\dots,0,\\
\pi+\arcsin(0.01),& n=1,2,\dots,
\end{array}
\right. 
\end{equation}
\begin{equation}
\Phi_\pi^3(n;0;0.1)=
\left\{
\begin{array}{llll}
0+\arcsin(0.1),& n=-1,-2,\dots\\
\pi-\arcsin(0.1),& n=-2,\\
2\pi+\arcsin(0.1),& n=-1,0,\\
\pi+\arcsin(0.1),& n=1,2,\dots,
\end{array}
\right.
\end{equation}
and
\begin{equation}
\Phi_\pi^3(n;0;0.55)= 
\left\{
\begin{array}{llll}
0+\arcsin(0.55),& n=-1,-2,\dots\\
\pi-\arcsin(0.55),& n=0,\\
\pi+\arcsin(0.55),& n=1,2,\dots.
\end{array}
\right.
\label{sf3055}
\end{equation}
One interesting point to note for the type~3 semikink is that the
number of sites with value~$2\pi$ is decreasing as~$\gamma$ increases.
Starting from the continuum approximation of a type~3 semikink as the
initial guess for the continuation program, the $2\pi$-plateau
disappears for $\gamma\geq\gamma^*(a)$~(see~(\ref{gs})). For
$\gamma>\gamma^*$ the configuration at $\eps=0$ is similar to the
stable type~1 $\pi$-kink~(\ref{lsf1}), apart from the value of the
phase at the site with $n=0$ (where the phase takes the value of an
unstable fixed point).

Because analytical calculation of the spectrum of the type 3 semikink has been obtained in the small coupling limit and bias current close to 1 (Eqs.\ (\ref{ell3})), it is worth comparing the analytical predictions with numerical computations.  
The theoretical calculations shows that for $\eps$ small and
  $\gamma$ close to~$1$, the type 3 semikink has at least two
  eigenvalues, one of which corresponds to a high frequency mode and
  the other to a positive eigenvalue.

  Using the continuation of~(\ref{sf3055}) for $\eps=0.25$, the
  spectrum of the type 3 lattice semikink is presented in
  Figure~\ref{specS3} as a function of the applied bias current
  $\gamma$. Our numerics show that when $\gamma$ is very close to
  $\gammacr$, the type~3 semikink has three eigenvalues, one of which
  corresponds to a high frequency mode and is below the phonon band,
  while the other two are above the phonon band. The birth of this
  high-frequency mode is shown in Figure~\ref{specS3}(c) and is
  qualitatively similar to the case of the type~1 lattice semikink.
  The two eigenvalues which exist for all values of $\gamma$ can be
  observed in Figure~\ref{specS3}(a) and (b).


\section{Conclusions}
\label{3sec7}

We have performed an existence and stability analysis for 
three types of lattice $\pi$-kink solutions of the discrete 
0-$\pi$ sine-Gordon equation and its continuum limits. 
Analytical results have been established in the continuum 
limits and in the weak-coupling case. It has been shown that 
in the continuous 0-$\pi$ sine-Gordon equation, $\pi$-kinks of 
type 1 are stable and the other types are unstable. The 
introduction of discreteness destabilizes the unstable 
$\pi$-kinks even more. An approximation to the largest 
eigenvalue of all types of $\pi$-kinks has been derived both 
in the continuum and the weak coupling limits.

For future research, it is of interest to study the nucleation 
of kinks and antikinks when a constant force, or bias current,
$\gamma$ that is above the critical value $\gamma_{\rm cr}$ is 
applied -- see Figure \ref{SFs_1}(b). One question that can be 
addressed is the mechanism and the frequency of the nucleation 
as a function of the applied constant force, especially in the 
presence of a damping coefficient (which has not been considered 
in this paper). In work in progress, the stability of the type 3 
semifluxons in the presence of defects in studied. These 
semifluxons are unstable, but the largest eigenvalue is close 
to zero. In fact, a type 3 semifluxon consists of a fluxon and 
a semifluxon with the opposite polarity. In experiments, the 
presence of a fluxon nearby a semifluxon can influence a 
junction measurement \cite{harl95}. Because a fluxon can be 
pinned by a defect \cite{kivs89}, one can expect to have a 
stable type 3 semifluxon when there is a defect present in 
the system. 

\section*{Acknowledgments}
H.S. wishes to thank Panayotis Kevrekidis for numerous useful interactions and discussions.



\begin{thebibliography}{99}

\bibitem {atki64} F. V. Atkinson, \emph{Discrete and Continuous Boundary Problems}, vol.\ 8 of Mathematics in Science and Engineering, (Academic Press, New York, 1964).
 
\bibitem {baes98} C. Baesens, S. Kim, and R. S. MacKay, \emph{Localised modes on localised equilibria}, Physica D {\bf113} (1998), pp. 242--247 
 
\bibitem {balm00} N. J. Balmforth, R. V. Craster, and P. G. Kevrekidis, 
\emph{Being stable and discrete},
Physica D {\bf 135} (2000), pp. 212--232.

\bibitem {base99} J.J.A. Baselmans, A.F. Morpurgo, B.J. van Wees, and T.M. Klapwijk,
\emph{Reversing the direction of the supercurrent in a controllable Josephson junction},
Nature {\bf 397} (1999), pp. 43--45.

\bibitem {brau98} O. M. Braun and Yu. S. Kivshar, 
\emph{Nonlinear dynamics of the Frenkel-Kontorova model},
Phys. Rep. {\bf 306} (1998), pp. 1--108.

\bibitem {brau97} O. M. Braun, Yu. S. Kivshar, and M. Peyrard, 
\emph{Kink's internal modes in the Frenkel-Kontorova model},
Phys. Rev. E {\bf 56} (1997), pp. 6050--6064.

\bibitem {bula77} L.N. Bulaevskii, V.V. Kuzii, and A. A. Sobyanin, 
\emph{Superconducting system with weak coupling to the current in the ground state},
JETP Lett. {\bf 25} (1977), pp. 290--294; 
L.N. Bulaevskii, V.V. Kuzii, A. A. Sobyanin, and P.N. Lebedev, 
\emph{On possibility of the spontaneous magnetic flux in a Josephson junction containing magnetic impurities},
Solid State Comm. {\bf 25} (1978), pp. 1053--1057.

\bibitem {cham00} A. Champneys and Yu. S. Kivshar, 
\emph{Origin of multikinks in dispersive nonlinear systems},
Phys. Rev. E {\bf 61} (2000), pp. 2551--2554.

\bibitem {chir79} B. V. Chirikov, \newblock\emph{A Universal Instability of Many-Dimensional Oscillator Systems}, 
Phys. Rep. {\bf52} (1979), pp. 264--379.

\bibitem{derk03} G. Derks, A. Doelman, S. A. van Gils and T. P. P. Visser,
\newblock
\emph{Travelling waves in a singularly perturbed sine-Gordon equation},
\newblock
Physica D {\bf 180} (2003), pp. 40--70.

\bibitem {guck86} J. Guckenheimer and P. Holmes, \emph{Nonlinear Oscillations, Dynamical Systems and Bifurcation of Vector Fields},
2nd ed. (Springer-Verlag, New York, 1986).

\bibitem{harl95} D. J. van Harlingen,
\emph{Phase-sensitive tests of the symmetry of the pairing state in the high-temperature superconductors--Evidence for $d_{x^2-y^2}$ symmetry},
Rev. Mod. Phys. {\bf 67} (1995), pp. 515--535.

\bibitem{henr81} D. Henry, {\it Geometry Theory of Semilinear Parabolic Equations},
Vol. 840 of {\it Lecture notes in mathematics} (Springer-Verlag, 1981).

\bibitem {hilg03} H. Hilgenkamp, Ariando, H. J. H. Smilde, D.H.A. Blank, G. Rijnders, H. Rogalla, J.R. Kirtley, and C.C. Tsuei, 
\emph{Ordering and manipulation of the magnetic moments in large-scale superconducting $\pi$-loop arrays},
Nature {\bf 422} (2003), pp. 50--53.

\bibitem{kato76} T.Kato, \emph{ Perturbation Theory of Linear Operators} (Springer, 1976).

\bibitem {kato97} T. Kato and M. Imada, 
\emph{Vortices and Quantum tunneling in Current-Biased 0-$\pi$-0 Josephson Junctions of $d$-wave Superconductors},
J. Phys. Soc. Jpn. {\bf 66} (1997), pp. 1445--1449.

\bibitem{kevr00} P. G. Kevrekidis and C. K. R. T. Jones, 
\emph{Bifurcation of internal solitary wave modes from the essential spectrum},
Phys. Rev. E {\bf 61} (2000), pp. 3114-3121.


\bibitem {kivs89} 
Yu. S. Kivshar and B. A. Malomed, 
\newblock
\emph{Dynamics of solitons in nearly integrable systems},
Rev. Mod. Phys. {\bf 61} (1989), pp. 763--915; {\it ibid.} {\bf 63} (1991), pp. 211 (Addendum).

\bibitem{kivs98} Yu. S. Kivshar, D. E. Pelinovsky, T. Cretegny, and M. Peyrard, 
\emph{Internal Modes of Solitary Waves},
Phys. Rev. Lett. {\bf 80} (1998), pp. 5032--5035.

\bibitem {kukl95} A. B. Kuklov, V.S. Boyko, and J. Malinsky, 
\emph{Instability in the current-biased 0-$\pi$ Josephson junction},
Phys. Rev. B {\bf 51} (1995), pp. 11965--11968; {\it ibid.} {\bf 55} (1997), pp. 11878 (Erratum).


\bibitem{mack95} R. S. MacKay and J. A. Sepulchre, 
\newblock
\emph{Multistability in networks of weakly coupled bistable units}, Physica D {\bf82} (1995), pp. 243--254.

\bibitem{mann97} 
 E.~Mann,
 \newblock
 \emph{Systematic perturbation theory for sine-Gordon solitons without use of
 inverse scattering methods},
 \newblock J.\ Phys.\ A: Math.\ Gen. \textbf{30} (1997), pp. 1227--1241.  

\bibitem{nomu96} Y. Nomura, Y. H. Ichikawa, and A. T. Filippov,\newblock
\emph{Stochasticity in the Josephson Map}, J. Plasma Phys. {\bf56} (1996), pp. 493–-506. 

\bibitem {peli05} D.~E.~Pelinovsky, P.~G.~Kevrekidis, and D.~J.~Frantzeskakis,\newblock
\emph{Stability of discrete solitons in nonlinear Schr\"{o}dinger lattices}, Physica D {\bf212} (2005), pp. 1--19.

\bibitem {peyr84} M. Peyrard and M. D. Kruskal, 
\emph{Kink dynamics in the highly discrete sine-Gordon system},
Physica D {\bf 14} (1984), pp. 88--102.

\bibitem {peyr82} M. Peyrard and M. Remoissenet, 
\emph{Solitonlike excitations in a one-dimensional atomic chain with a nonlinear deformable substrate potential},
Phys. Rev. B {\bf 26} (1982), pp. 2886--2899.

\bibitem {quin00} N. R. Quintero, A. Sánchez, and F. G. Mertens, 
\emph{Anomalous Resonance Phenomena of Solitary Waves with Internal Modes},
Phys. Rev. Lett. {\bf 84} (2000), pp. 871--874.

\bibitem {rose03} P. Rosenau, \emph{Hamiltonian dynamics of dense chains and lattices: or how to correct the continuum}, 
Phys. Lett. A {\bf311} (2003), pp. 39--52 

\bibitem {ryaz01} V.V. Ryazanov, V.A. Oboznov, A.Yu. Rusanov, A.V. Veretennikov, A.A. Golubov, and J. Aarts, 
\emph{Coupling of Two Superconductors through a Ferromagnet: Evidence for a $\pi$ Junction},
Phys. Rev Lett. {\bf 86} (2001), pp. 2427--2430.

\bibitem{scot99} A. Scott, {\it Nonlinear science: emergence and dynamics of coherent structures}, Oxford University Press, 1999.

\bibitem{susa04} H. Susanto and S. A. van Gils, 
\emph{Instability of a lattice semifluxon in a current-biased 0-$\pi$ array of Josephson junctions},
Phys. Rev. B {\bf 69} (2004), pp. 092507--092510.


\bibitem{susa03} 
\newblock
H. Susanto, S. A. van Gils, T. P. P. Visser, Ariando, H. J. H. Smilde, and H. Hilgenkamp, 
\emph{Static semifluxons in a long Josephson junction with $\pi$-discontinuity points},
Phys. Rev. B. {\bf68} (2003), pp. 104501--104508.

\bibitem{Titch}
E.C.~Titchmarsh,
\newblock \emph{Eigenfunction expansions associated with second-order
  differential equations} (2nd edition),
\newblock Oxford University Press, 1962.

\bibitem {tsue00} C.C. Tsuei and J.R. Kirtley, 
\emph{Pairing symmetry in cuprate superconductors},
Rev. Mod. Phys. {\bf 72} (2000), pp. 969--1016.

\bibitem {vav06} O. V\'avra, S. Ga\v{z}i, D. S. Golubovi\'c, I. V\'avra, J. D\'erer, J. Verbeeck, G. Van Tendeloo, and V. V. Moshchalkov, 
\emph{The 0 and the $\pi$ phase Josephson coupling through an insulating barrier with magnetic impurities}, Phys. Rev. B {\bf74} (2006), pp. 020502--020505.

\bibitem{yosida95} K. Yosida, \emph{Functional Analysis} (Springer, 1995).

\end{thebibliography}
\end{document}